\documentclass[prd,aps,twocolumn,a4paper,nofootinbib]{revtex4}
\usepackage{amsmath}
\usepackage{amsfonts}
\usepackage{amssymb}	
\usepackage{graphicx}
\usepackage{bm}
\usepackage{color}
\usepackage{ulem}
\usepackage{commath}
\allowdisplaybreaks

\def\GMc2{G M_{\odot} c^{-2}}

\def\lm{{\ell m}}

\def\lm{{\ell m}}

\def\lm{{\ell m}}

\def\F{{\cal F}}

\newcommand\be{\begin{equation}}
\newcommand\ee{\end{equation}}

\def\TEOBResumS{\texttt{TEOBResumS}}

\newcommand{\gE}{{\gamma_\textrm{E}}}

\begin{document}
\title{The next generation: Impact of high-order analytical information on effective one body waveform models for 
       noncircularized, spin-aligned black hole binaries.}
\author{Alessandro \surname{Nagar}${}^{1,2}$}
\author{Piero \surname{Rettegno}${}^{1,3}$}
\affiliation{${}^1$INFN Sezione di Torino, Via P. Giuria 1, 10125 Torino, Italy}
\affiliation{${}^2$Institut des Hautes Etudes Scientifiques, 91440 Bures-sur-Yvette, France}
\affiliation{${}^{3}$ Dipartimento di Fisica, Universit\`a di Torino, via P. Giuria 1, 10125 Torino, Italy}

\begin{abstract}
We explore the performance of an updated effective-one-body (EOB) model for spin-aligned coalescing 
black hole binaries designed to deal with any orbital configuration. The model stems  from previous work 
involving the \TEOBResumS{} waveform model, but incorporates recently computed analytical information up 
to fifth post-Newtonian (PN) order in the EOB potentials. The dynamics is then informed by  
Numerical Relativity (NR) quasi-circular simulations (incorporating also recently computed 4PN spin-spin and,
optionally, 4.5PN spin-orbit terms). The so-constructed model(s) are then compared  to various kind of 
NR simulations, covering either quasi-circular inspirals, eccentric inspirals and scattering configurations. 
For quasi-circular (534 datasets) and eccentric (28 datasets) inspirals up to coalescence, the EOB/NR unfaithfulness is 
well below $1\%$ except for a few outliers in the high, positive, spin corner of the parameter space, 
where however it does not exceed the $3\%$ level. 
The EOB values of the scattering angle are found to agree ($\lesssim 1\%$) with the NR predictions for most configurations, 
with the largest disagreement of only $\sim 4\%$ for the most relativistic one. The inclusion of some 
high-order analytical information  in the orbital sector is useful to improve the EOB/NR agreement 
with respect to previous work, although the use of NR-informed functions is still crucial to accurately 
describe the strong-field dynamics and waveform.
\end{abstract}
   
\date{\today}

\maketitle

\section{Introduction}
\label{sec:intro}
There is an ongoing effort in building up accurate waveform model for {\it noncircularized}
coalescing black-hole binaries~\cite{Chiaramello:2020ehz,Nagar:2020xsk,Islam:2021mha,Nagar:2021gss,Albanesi:2021rby,Liu:2021pkr,Khalil:2021txt}. 
In particular, the generalization of the quasicircular models \TEOBResumS{}~\cite{Nagar:2020pcj,Riemenschneider:2021ppj} to 
nonquasicircular configurations~\cite{Chiaramello:2020ehz,Nagar:2020xsk,Nagar:2021gss} 
has allowed the construction of the first, and currently only, effective-one-body waveform 
model for spin-aligned black hole binaries that is able to accurately deal 
with both hyperbolic captures~\cite{East:2012xq,Gold:2012tk,Nagar:2020xsk,Nagar:2021gss}
and eccentric inspirals~\cite{Chiaramello:2020ehz,Nagar:2021gss} 
(see also~\cite{Cao:2017ndf,Liu:2019jpg,Liu:2021pkr,Yun:2021jnh} for a different EOB-based construction limited to eccentric inspiral).
 In particular, the model of Ref.~\cite{Chiaramello:2020ehz,Nagar:2021gss}
has been used to analyze the GW source 
GW190521~\cite{LIGOScientific:2020iuh,LIGOScientific:2020ufj} under the hypothesis that it is the result 
of an hyperbolic capture~\cite{Gamba:2021gap}. However, the most recent results of 
Ref.~\cite{Nagar:2021gss} should be further improved, since the quasicircular limit 
of the model is considerably {\it less accurate} (EOB/NR unfaithfulness $\simeq 1\%$) 
than the native quasi-circular  model \TEOBResumS{}~\cite{Nagar:2020pcj,Riemenschneider:2021ppj}
(EOB/NR unfaithfulness $\simeq 0.1\%$). The purpose of this paper is to show that certain 
modifications to the underlying analytical structure of the EOB dynamics allow to do so 
and thus obtain a waveform model that: (i) is highly NR faithful for quasicircular
coalescing BBHs; (ii) improves the EOB/NR agreement for the limited number of eccentric NR inspiral waveforms
currently publicly available; (iii) it similarly allows for an improved agreement between EOB and NR scattering angles.
Technically, this is accomplished using the model of Ref.~\cite{Nagar:2021gss} where some of the analytical building 
blocks of the Hamiltonian are modified. In particular: (i) we implement recently computed 5PN-accurate
information~\cite{Bini:2019nra,Bini:2020wpo} in the EOB potentials $(D,Q)$; (ii) the potentials $(A,D)$ are resummed
using diagonal ($P^3_3$ for $A$) and near diagonal ($P^3_2$ for $D$) Pad\'e approximant instead of the
$P^1_5$ and $P^0_3$ approximants that have been shared by all realizations of \TEOBResumS{} up to now.
In addition, we also explore the impact of new analytical information in the spin sector. In particular, for what
concerns spin-spin interaction, we incorporate all available information up to 4PN (NNLO)~\cite{Levi:2015ixa,Levi:2016ofk}, 
following the EOB implementation of Ref.~\cite{Nagar:2018plt}. Similarly, for the spin-orbit sector we also investigate 
the effect of the next-to-next-to-next-to-leading order (N$^3$LO) contribution recently obtained in Refs.~\cite{Antonelli:2020aeb,Antonelli:2020ybz}.

The paper is organized as follows. In Sec.~\ref{sec:teobresums} we review the analytical elements of
the EOB waveform model we are introducing here, in particular highlighting the differences with previous
works. Section~\ref{sec:qc_dynamics} illustrates the quasi-circular limit of the model, how it is informed by NR
simulations and how it performs on the SXS waveform catalog. Similarly, Sec.~\ref{sec:ecc_dynamics} reports 
EOB/NR comparisons for eccentric inspirals, while Sec.~\ref{sec:scattering} focuses on the scattering angle.
The paper is ended by concluding remarks in Sec.~\ref{sec:conclusions}, while Appendix~\ref{sec:barF_NBR}
presents some updates and corrections to the findings of Ref.~\cite{Nagar:2021gss}.
If not otherwise specified, we use units with $c=G=1$.

\section{EOB dynamics with 5PN terms}
\label{sec:teobresums}

\subsection{The EOB potentials}
\label{sec:Ham}
The structure of the dynamics and waveform of the EOB eccentric model discussed
here is the same as Ref.~\cite{Nagar:2021gss} except for the PN accuracy of the potentials
$(A,D,Q)$ and their resummed representation. Before giving details about them, 
let us recall the basic notation adopted.
We use mass-reduced phase-space variables $(r,\varphi,p_\varphi,p_{r_*})$,  related to the physical 
ones by $r=R/M$ (relative separation), $p_{r_*}=P_{R_*}/\mu$ (radial momentum), $\varphi$ (orbital phase),
$p_\varphi=P_\varphi/(\mu M)$ (angular momentum) and $t=T/M$ (time),
where $\mu\equiv m_1 m_2/M$ and $M\equiv m_1+m_2$. 
The radial momentum is 
$p_{r_*}\equiv (A/B)^{1/2}p_r$, where $A$ and $B$ are the EOB potentials 
(with included spin-spin interactions, see below~\cite{Damour:2014sva}) 
and $D=A B$ (for nonspinning systems).
The EOB Hamiltonian is $\hat{H}_{\rm EOB}\equiv H_{\rm EOB}/\mu=\nu^{-1}\sqrt{1+2\nu(\hat{H}_{\rm eff}-1)}$, 
with $\nu\equiv \mu/M$ and $\hat{H}_{\rm eff}=\tilde{G}p_\varphi + \hat{H}^{\rm orb}_{\rm eff}$, 
where $\tilde{G}p_\varphi$ incorporates odd-in-spin (spin-orbit) effects while 
$\hat{H}^{\rm orb}_{\rm eff}$ takes into account even-in-spin effects through the use of 
the centrifugal radius $r_c$~\cite{Nagar:2018zoe}, that we discuss below.
The orbital Hamiltonian for non-spinning systems reads
\be
\hat{H}_{\rm orb}=\sqrt{A(u)(1+p_\varphi u^2) + p_{r_*}^2 + Q(u,p_{r_*})},
\ee
where $u\equiv 1/r$. The Taylor expanded expressions of the $(A,D)$ potential up to 5PN accuracy read
\begin{widetext}
\begin{align}
A_{\rm 5PN}(u) &= 1 - 2 u + 2 \nu \,u^3 + \nu \left(\frac{94}{3}-\frac{41\pi^2}{32}\right) u^4 +\nonumber\\
&\quad+ \left[
\left( \frac{2275\pi^2}{512} - \frac{4237}{60} + \frac{128}{5}\gE + \frac{256}{5}\ln(2) \right) \nu + \left( \frac{41\pi^2}{32} - \frac{221}{6} \right)\nu^2
+ \frac{64}{5}\nu\,\ln(u) \right] u^5 + \nonumber \\
&\quad+ \nu \left[a_{6}^c  + \left(-\frac{7004}{105} - \frac{144}{5}\nu\right)\,\ln(u)\right]u^6,
\\
\label{eq:D}
D_{\rm 5PN}(u) &= 1 - 6\nu\,u^2 - \left(52\nu - 6\nu^2\right) u^3 +
\nonumber\\
&\quad + \left[
\left( \frac{533}{45} + \frac{23761\pi^2}{1536} - \frac{1184}{15}\gE +\frac{6496}{15}\ln(2) - \frac{2916}{5}\ln(3) \right)\nu
+ \left(-\frac{123\pi^2}{16} + 296\right)\nu^2 - \frac{592}{15}\nu\ln(u) \right] u^4 + \nonumber \\
&\quad + \Bigg[\left(-\frac{294464}{175}+\frac{63707 \pi ^2}{512}+\frac{2840}{7}\gE-\frac{120648}{35}\ln(2)+\frac{19683}{7}\ln(3)\right)\nu+ \nonumber \\
&\quad\quad\quad+\left(-d_5^{\nu^2}-\frac{2216}{105}+\frac{6784}{15}\gE+\frac{326656}{21}\ln(2)-\frac{58320}{7}\ln(3)\right)\nu^2+\nonumber \\
&\quad\quad\quad+\left(-\frac{1285}{3}+\frac{205 \pi^2}{16}\right) \nu^3+\left(\frac{1420}{7}\nu+\frac{3392}{15}\nu^2\right)\ln(u) \Bigg]u^5,
\end{align}
\end{widetext}
where $\gamma_E = 0.577216\dots$ and we kept implicit the 
coefficient $a_6^c(\nu)$. Its analytically known expression reads~\cite{Bini:2013zaa,Damour:2014jta,Damour:2015isa,Damour:2016abl}
\begin{align}
\label{eq:a6c_anlyt}
&a_{\rm 6 \, anlyt}^c(\nu) = -\frac{1066621}{1575} + \frac{246367 \pi ^2}{3072} \nonumber\\
& - \frac{14008}{105}\gE -\frac{31736}{105}\ln(2) + \frac{243}{7}\ln(3) \\
&+ \left(\frac{64}{5}-\frac{288}{5}\gE+\frac{928}{35}\ln(2)-\frac{972}{7}\ln(3) + a_6^{\nu^2}\right) \nu +4 \nu^2.\nonumber
\end{align}
Note that both Eq.~\eqref{eq:D} and~\eqref{eq:a6c_anlyt} present two yet undetermined 
analytical coefficients, $(a_6^{\nu^2},d_5^{\nu^2})$. For simplicity, in this work we impose 
$d_5^{\nu^2}=0$.  By contrast, following previous works, we will not use the analytical 
expression $a_{\rm 6 \, anlyt}^c(\nu)$, but rather consider $a_6^c(\nu)$ as an undetermined 
function of $\nu$ that is informed using NR simulations. The differences between the resulting
NR-informed $A$ function and the one that uses  $a_{\rm 6 \, anlyt}^c(\nu)$  will be discussed below.
The $Q$ function at 5PN accuracy was obtained in Ref.~\cite{Bini:2020wpo}. For simplicity, here we only consider the {\it local}
part of $Q$ at 5PN\footnote{We have also attempted to incorporate the nonlocal part, but the expression is rather complicate
and it seems to degrade the robustness of the model in strong field. Since its effects deserve a more detailed study we 
defer it to future work.}. Once $p_r$ is rewritten in terms of $p_{r_*}$, the function reads
\begin{widetext}
\begin{align}
Q_{\rm 5PNloc}(u, p_{r_*}) &= 2(4-3\nu )\nu\,u^2p_{r_*}^4
+ \left[ \left( -\frac{4348}{15} + \frac{496256}{45}\ln(2) - \frac{33048}{5}\ln(3) \right) \nu
- 131 \nu^2 + 10 \nu^3 \right] u^3 p_{r_*}^4
\nonumber\\
&\quad
+ \left[\left(-\frac{827}{3} - \frac{2358912}{25}\ln(2) + \frac{1399437}{50}\ln(3)
+ \frac{390625}{18}\ln(5) \right)\nu - \frac{27}{5}\nu^2 + 6\nu^3 \right]u^2\,p_{r_*}^6 + \nonumber \\
&\quad+\left[\left(-\frac{32957}{10}-\frac{28306944}{25}\ln(2)+\frac{8396622}{25}\ln(3)+\frac{781250}{3}\ln(5)\right)\nu -\frac{393}{5}\nu^2 + 188 \nu^3 -14 \nu^4\right]u^3 p_{r_*}^6 +\nonumber \\
&\quad+ \Bigg[\left(-\frac{6328799}{3150}-\frac{93031 \pi ^2}{1536}+\frac{3970048}{45}\ln(2)-\frac{264384}{5}\ln(3)\right)\nu +\left(-\frac{5075}{3}+\frac{31633 \pi
	^2}{512}\right) \nu ^2 \nonumber \\
&\quad\quad\quad+\left(792-\frac{615 \pi ^2}{32}\right) \nu ^3\Bigg]u^4 p_{r_*}^4+\left(\frac{6}{7}\nu+\frac{18}{7}\nu^2+\frac{24}{7}\nu^3-6 \nu ^4\right)u^2 p_{r_*}^8.
\end{align}
\end{widetext}
Here we will keep the function $Q$ in its PN-expanded form. By contrast, both the $(A,D)$ functions will
be resummed using Pad\'e approximants, although  with different choices with respect to previous work.
Within the \TEOBResumS{} models, the formal 5PN-accurate $A$ function is always 
resummed via a $(1,5)$ Pad\'e approximant. As we will illustrate below, this approximant 
develops a spurious pole when  $a_6^c=a_{\rm 6 anlyt}^c$. Since we also want to get a 
handle on the performance of the pure analytical information, we  are forced 
to change the resummation choice. To do so, 
we follow the most straightforward approach and use the diagonal Pad\'e approximant, that is
\be
A(u,\nu;a_6^c)=P^{3}_{3}[A_{\rm 5PN}(u,\nu;a_6^c)],
\ee
where it is intended that the $\ln(u)$ terms are treated as numerical constants when computing the Pad\'e.
\begin{figure}[t]
	\center
\includegraphics[width=0.42\textwidth]{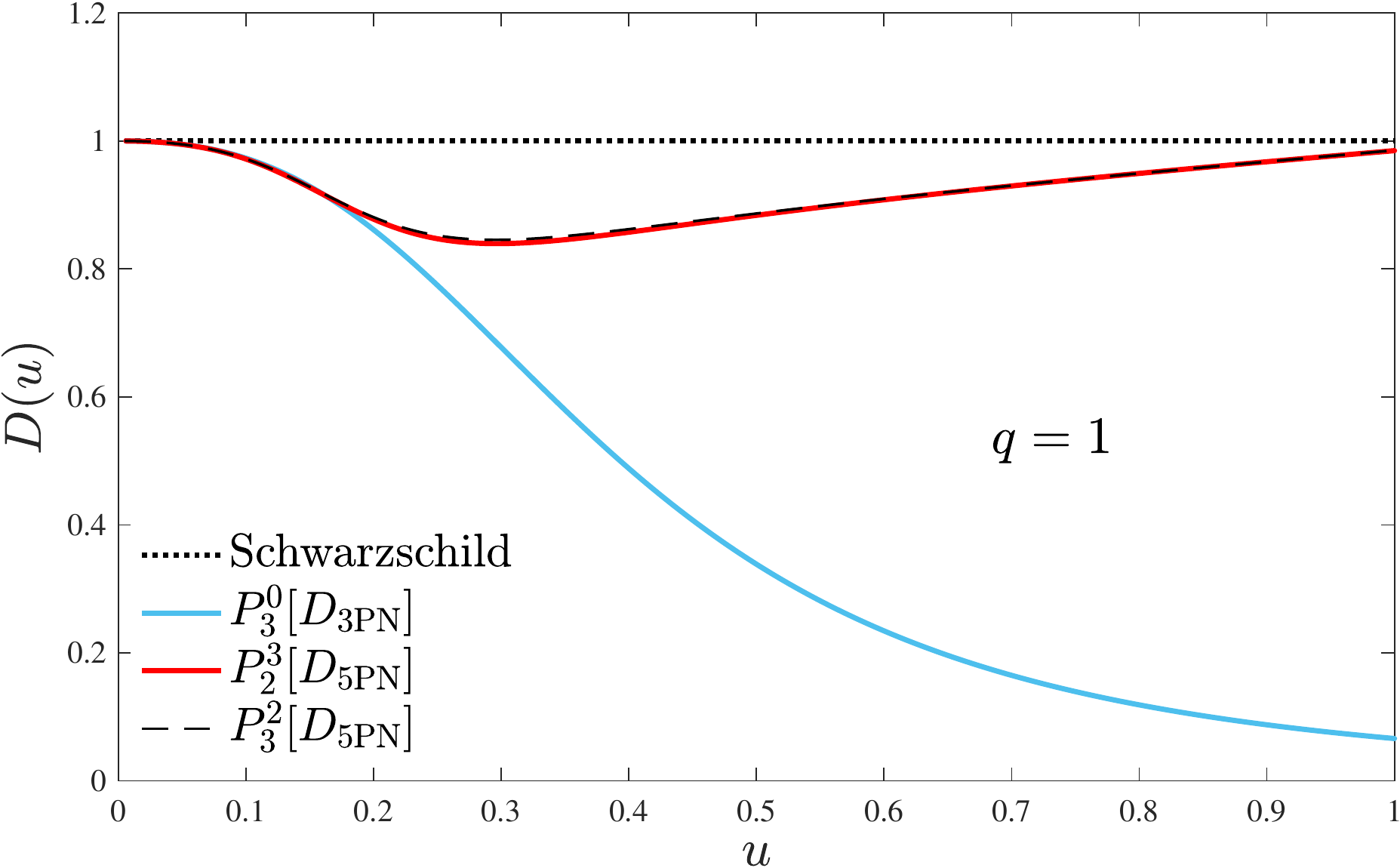}
\caption{\label{fig:Dold} 
	Comparison between different realizations of the $D$ function for $q=1$: the $P^0_3[D_{\rm 3PN}]$ used in the standard implementation 
	of \TEOBResumS{} (blue online). Since the  $P^0_5[D_{\rm 5PN}]$ develops a spurious pole for $0<u<1$, we plot instead
	$P^3_2[D_{\rm 5PN}]$ and $P^2_3[D_{\rm 5PN}]$, that are almost identical. However, the latter develops a spurious pole
	well outside the  domain (around $u\sim 8$). Note that both approximants are quantitatively consistent with the Schwarzschild
	potential,  $D = 1$ (dotted  line). We choose the $P^3_2[D_{\rm 5PN}]$ Pad\'e approximant to represent the $D$ function.}
\end{figure}
The 3PN-accurate $D$ function, that in \TEOBResumS{}, is resummed using a $(0,3)$ approximant.
When the same is attempted with the 5PN-accurate function (with $d_5^{\nu^2}=0$), spurious
poles again show up for any value of $\nu$. 
By contrast, the quasi-diagonal Pad\'e approximants $P^3_2$ and $P^2_3$ stabilize the series:
they are very similar to each other and generally consistent with the Schwarzschild value, 
$D_{\rm Schw} = 1$ . Figure~\ref{fig:Dold} highlights these facts for the case $q=1$.
Eventually, we choose to resum $D$ as
\begin{align}
D &=P^3_2[D_{\rm 5PN}] ,
\end{align}
because the $P^2_3$ develops a spurious pole for large 
(even though unphysical, $u\sim 8$) values of $\nu$.
By contrast, for simplicity we use $Q$ in its PN-expanded form-

\subsection{The spin sector}
\label{sec:spin}

When taking into account spinning bodies, the radial variable $r$ is replaced,
within $\hat{H}_{\rm eff}^{\rm orb}$, by the centrifugal radius $r_c$ that is used 
to incorporate spin-spin terms~\cite{Damour:2014sva}. Its explicit expression,
that includes spin-spin terms up to NNLO, will be described in detail in Sec.~\ref{sec:rc_nnlo}
below. Here, let us just recall that we define $u_c \equiv r_c^{-1}$ and that the $D$
function in the spinning case is defined as
\be
D\equiv \dfrac{r^2}{r_c^2}D_{\rm orb}(u_c),
\ee
where 
\be
D_{\rm orb}=P^3_2[D_{\rm 5PN}(u_c)],
\ee
using $D_{\rm 5PN}$ from Eq.~\eqref{eq:D}. As mentioned above, the effective 
Hamiltonian with spin-orbit couplings is written as
\be
\hat{H}_{\rm eff}= \hat{H}_{\rm eff}^{\rm orb} + \tilde{G}p_\varphi,
\ee
with
\be
\tilde{G}\equiv \left( G_S \hat{S} + G_{S_*} \hat{S}_* \right),
\ee
where we are using
\begin{align}
\hat{S}  &\equiv  (S_1 + S_2) M^{-2} ,\\
\hat{S}_*&\equiv \left(\dfrac{m_2}{m_1}S_1 + \dfrac{m_1}{m_2}S_2\right) M^{-2} ,
\end{align}
and the gyro-gravitomagnetic functions are factorized as 
\begin{align}
G_S &= G_S^0 \hat{G}_S, \\
G_{S_*} &= G_{S_*}^0 \hat{G}_{S_*}.
\end{align}
Here, the leading-order contributions read
\begin{align}
G_S^0 &= 2uu_c^2 \ ,\\
G_{S_*}^0 &= \frac32 u_c^2 \ ,
\end{align}
while the higher PN corrections are formally including
up to N$^3$LO, corresponding to 4.5PN order in the following
resummed form
\begin{widetext}
\begin{align}
\label{eq:GS}
\hat{G}_S &= (1 + c_{10}u_c + c_{20}u_c^2 + c_{30}u_c^3 + c_{40} u_c^4 + c_{02}p_{r^*}^2 + c_{12}u_c p_{r^*}^2 + c_{04}p_{r^*}^4+c_{22}p_{r_*}^2 u_c^2 + c_{14} u_c p_{r_*}^4 + c_{06}p_{r_*}^6)^{-1} ,\\
\label{eq:GSs}
\hat{G}_{S_*} & = (1 + c^*_{10}u_c + c^*_{20}u_c^2 + c^*_{30}u_c^3 +c^*_{40} u_c^4+ c^*_{02}p_{r^*}^2 + c^*_{12}u_c p_{r^*}^2 + c^*_{04}p_{r^*}^4+c_{22}^*p_{r_*}^2 u_c^2 + c_{14}^* u_c p_{r_*}^4 + c_{06}^*p_{r_*}^6)^{-1},
\end{align}
\end{widetext}
where however we also have included two coefficients $(c_{40},c^*_{40})$ that belong
to the next-to-next-to-next-to-next-to-leading (N$^4$LO) order. 
Following previous work, we fix $c_{40}=0$ and $c^*_{40} = 2835/256$.
The second value comes from the expansion of the Hamiltonian of a spinning particle around
a spinning black hole~\cite{Damour:2014sva}. Within this gauge, specifying the 
N$^3$LO spin-orbit contribution is equivalent to specifying $8$ numerical coefficients. Here we consider two
separate options: (i) on the one hand, we use a N$^3$LO parametrization that is tuned to
NR simulations, following the usual procedure adopted within the \TEOBResumS{}
model; (ii) on the other hand, we also consider an {\it analytical version} of the
N$^3$LO contribution that has been recently obtained with a mixture of several
analytical techniques~\cite{Antonelli:2020aeb,Antonelli:2020ybz}.

\subsubsection{NR-informed spin-orbit description}
\label{SO:nr}
Following previous work~\cite{Damour:2014sva}, at N$^3$LO order we only consider
\begin{align}
c_{30} &=\nu c_3, \\
c^*_{30} &= \nu c_3 + \frac{135}{32} ,
\end{align}
where $c_3$ is the NR-informed tunable parameter, while all other N$^3$LO coefficients are fixed 
to zero $c_{22}=c_{14}=c_{06}=c_{22}^*=c_{04}^*=c_{06}^*=0$. The NR-informed expression of $c_3$,
that will be found to be a function of $\nu$ and of the spins, will be discussed in Sec.~\ref{sec:nr_inform}
below.

\subsubsection{Fully analytical spin-orbit description}
\label{SO:anlyt}
Recently, Refs.~\cite{Antonelli:2020aeb,Antonelli:2020ybz} used first-order self-force
(linear-in-mass-ratio) results to obtain arbitrary-mass-ratio results for the N$^3$LO correction
to the spin-orbit sector of the Hamiltonian. The N$^3$LO contribution is given by Eqs.(8) and (9) 
of Ref.~\cite{Antonelli:2020aeb}. Once incorporated within the expression of $(G_{S},G_{S_*})$ 
of Eqs.~\eqref{eq:GS}-\eqref{eq:GSs} above, the explicit expressions of the N$^3$LO coefficients read
\begin{align}
c_{30} &=  \left(\dfrac{80399}{2304} - \dfrac{241}{384}\pi^2\right)\nu - \dfrac{31}{16}\nu^2 + \dfrac{397}{4096}\nu^3,\\
c_{22} &=  \dfrac{10563}{128}\nu - \dfrac{2273}{64}\nu^2 - \dfrac{2999}{4096}\nu^3,\\
c_{14} &=  -\dfrac{1421}{256}\nu + \dfrac{1257}{128}\nu^2-\dfrac{2201}{4096}\nu^3,\\
c_{06} &=  -\dfrac{7}{256}\nu - \dfrac{9}{128}\nu^2 + \dfrac{83}{4096}\nu^3,\\
c^*_{30} &=  \dfrac{135}{32} + \left(\dfrac{5501}{144} - \dfrac{41}{48}\pi^2\right)\nu - \dfrac{5}{32}\nu^2 + \dfrac{5}{16}\nu^3, \\
c^*_{22} & =  \dfrac{773}{64} + \dfrac{2313}{32}\nu - \dfrac{245}{8}\nu^2 - 2\nu^3,\\
c^*_{14} &=  \dfrac{35}{48} +\dfrac{115}{6}\nu + \dfrac{395}{96}\nu^2 - \dfrac{9}{16}\nu^3,\\
c^*_{06} & = -\dfrac{5}{96} + \dfrac{5}{16}\nu + \dfrac{37}{32}\nu^2 - \dfrac{\nu^3}{16} \ .
\end{align}
\subsubsection{Spin-spin effects: NNLO accuracy}
\label{sec:rc_nnlo}
The spin-spin sector incorporates NNLO information~\cite{Levi:2015ixa,Levi:2016ofk} within the centrifugal
radius $r_c$, according to the usual scheme typical of the \TEOBResumS{} 
Hamiltonian~\cite{Damour:2014sva}. In particular, we use here the analytical expressions obtained in 
Ref.~\cite{Nagar:2018plt} once specified to the BBH case. However, to robustly incorporate NNLO 
information in strong field, it is necessary to implement it in resummed form. 
To start with, we formally factorized the centrifugal radius as
\be
\label{eq:rc2}
r_c^2 = (r^{\rm LO}_c)^2 \hat{r}_c^2,
\ee
where the $(r^{\rm LO}_c)^2$ is the LO contribution,
$\hat{r}_c^2$ the PN corrections up to NNLO. Concretely, we have
\be
(r_c^{\rm LO})^2 = r^2 + \tilde{a}_0^2\left(1+\dfrac{2}{r}\right) \ ,
\ee
where
\be
\tilde{a}_0 = X_1 \chi_1 + X_2 \chi_2
\ee
with $\chi_i\equiv S_i/m_i^2$, with $i=1,2$, and $X_i \equiv m_i/M$ (with $X_1\geq X_2$), while
$\hat{r}_c^2$ explicitly reads
\be
\label{eq:hat_rc2}
\hat{r}_c^2 = 1 + \dfrac{\delta a^2_{\rm NLO}}{r (r_c^{\rm LO})^{2}} + \dfrac{\delta a ^2_{\rm NNLO}}{r^2 (r_c^{\rm LO})^{2}},
\ee
where we have [see Eqs.~(19) and (20) of Ref.~\cite{Nagar:2018plt}]
\begin{align}
\delta a^2_{\rm NLO} &= -\dfrac{9}{8}\tilde{a}_0^2 -\dfrac{1}{8}(1+4\nu)\tilde{a}_{12}^2 + \dfrac{5}{4} X_{12}\tilde{a}_0\tilde{a}_{12}, \\
\delta a^2_{\rm NNLO} & =  -\left(\dfrac{189}{32} + \dfrac{417}{32}\nu\right)\tilde{a}_0^2 \nonumber\\
                                    & + \left(\dfrac{11}{32} - \dfrac{127}{32}\nu + \dfrac{3}{8}\nu^2\right)\tilde{a}_{12}^2 \nonumber\\ 
                                    &+ \left(\dfrac{89}{16} - \dfrac{21}{8}\nu\right) X_{12}\tilde{a}_0\tilde{a}_{12} \ ,
\end{align}
where $X_{12}\equiv X_1-X_2$ and $\tilde{a}_{12}\equiv \tilde{a}_1-\tilde{a}_2$.
Direct inspection of the Taylor-expanded expression of $r_c^2$ shows its oscillatory behavior when
moving from LO to NNLO. This suggests that to fruitfully incorporate the NNLO term, some resummation
procedure should be implemented. To do so, we simply note that $\hat{r}_c^2$ given by Eq.~\eqref{eq:hat_rc2} 
has the structure $1+ c_{\rm NLO}\epsilon + c_{\rm NNLO}\epsilon^2$, where $\epsilon$ is a formal 
PN ordering parameter\footnote{Note that NNLO spin-spin effect correspond to 4PN accuracy, while 
the LO is 2PN accuracy~\cite{Levi:2015ixa}. So, when LO is factored out one is left with a residual
expansion that is 2PN accurate.}, and it can be robustly resummed taking a $P^0_2$ approximant in $\epsilon$. 
From now on, it is thus intended that we will work with the Pad\'e resummed 
quantity $P^0_2[\hat{r}_c^2,\epsilon]$ instead of $\hat{r}_c^2$ in Taylor-expanded form.

\subsection{Radiation reaction and waveform}
\label{sec:RR}
The prescription for the radiation reaction force we are using 
follows  Ref.~\cite{Nagar:2021gss} (see also~\cite{Albanesi:2021rby}), 
although minimal details about the structure of $\hat{\F}_\varphi$ 
were explicitly reported there. We complement here the discussion of~\cite{Nagar:2021gss} 
for clarity and completeness. The global structure of  $\hat{F}_\varphi$ 
is that of the quasi-circular version of \TEOBResumS{},  as discussed in 
Ref.~\cite{Damour:2014sva}. In particular, its formal expression reads
\be
\hat{\F}_\varphi = -\dfrac{32}{5} \nu \, r_\omega^4 \, \Omega^5 \hat{f}(\Omega),
\ee
where $r_\omega$ is given by Eq.~(70) of Ref.~\cite{Damour:2014sva}, $\Omega=\dot{\varphi}$
is the orbital frequency and $\hat{f}(\Omega)$ is the Newton-normalized flux function. 
For the quasi-circular model $\hat{f}(\Omega)$ is the circular flux function given by the sum of several
modes $\hat{f}_\lm$ where the hat indicates that each $(\ell,m)$ multipole is normalized 
by the $\ell=m=2$ Newtonian flux $F_{22}^{\rm Newt}=32/5\,\nu \, \Omega^{10/3}$. 
Each circularized multipole is then factorized and resummed according
to Ref.~\cite{Nagar:2020pcj}. In the most general case of motion along noncircular orbits, 
each Newton-normalized multipoles acquires a noncircular factor, so that the flux 
can be formally written as
\be
\hat{f}(\Omega)=\sum_{\ell =2}^{8}\sum_{m=-\ell}^{\ell}\hat{f}_{\ell m}\hat{f}_{\ell m}^{\rm non-circular} . 
\ee
Here we will consider only $\hat{f}_{22}^{\rm non-circular}\neq 0$ and use it in its 
Newtonian approximation, see Ref.~\cite{Chiaramello:2020ehz}. 
The Newtonian noncircular factor reads
\begin{align}
\label{eq:Fnewt}
&\hat{f}_{22}^{\rm non-circular}=\hat{f}_{22}^{\rm {Newt}_{nc}}=1+\dfrac{3}{4}\dfrac{\ddot{r}^2}{r^2\Omega^4} - \dfrac{\ddot{\Omega}}{4\Omega^3}
+\dfrac{3\dot{r}\dot{\Omega}}{r\Omega^3} \nonumber\\
& +\dfrac{4\dot{r}^2}{r^2\Omega^2}
+\dfrac{\ddot{\Omega}\dot{r}^2}{8 r^2\Omega^5} + \dfrac{3}{4}\dfrac{\dot{r}^3\dot{\Omega}}{ r^3\Omega^5}
+\dfrac{3}{4}\dfrac{\dot{r}^4}{r^4\Omega^4} + \dfrac{3}{4}\dfrac{\dot{\Omega}^2}{\Omega^4}\\
&-\dddot{r}\left(\dfrac{\dot{r}}{2r^2\Omega^4}+\dfrac{\dot{\Omega}}{8r\Omega^5}\right)
+\ddot{r}\left(-\dfrac{2}{r\Omega^2}+\dfrac{\ddot{\Omega}}{8r \Omega^5}+\dfrac{3}{8}\dfrac{\dot{r}\dot{\Omega}}{r^2\Omega^5}\right)\nonumber.
\end{align}
For what concerns the $\ell=m=2$ waveform, everything follows Ref.~\cite{Nagar:2020pcj}
except for a change in one of the functions that determine the next-to-quasi-circular (NQC)
correction to the amplitude. In particular, it turns out that the function 
$n_2^{22}=\underline{\ddot{r}}^{(0)}/(r\Omega^2)$, where $\underline{\ddot{r}}^{(0)}$ 
is an approximation to the second derivative of the radial separation, given by 
Eq.~(3.37) of Ref.~\cite{Nagar:2019wds}, is not robust in strong field in conjunction 
with the new EOB potentials. As an alternative, we use instead
\be
n^{22}_2 = n^{22}_1 (p_{r_*})^2 \ ,
\ee
where
\be
n_1^{22} =\left(\dfrac{p_{r_*}}{r\Omega}\right)^2.
\ee
These choices ensure the construction of the NR-informed amplitude around merger 
that is robust, although it might sometimes slightly overestimate (by a few percents) 
the corresponding NR one.

\section{Quasi-circular configurations}
\label{sec:qc_dynamics}

\subsection{Effective one body dynamics informed by NR simulations}
\label{sec:nr_inform}
\begin{figure}[t]
\center
\includegraphics[width=0.45\textwidth]{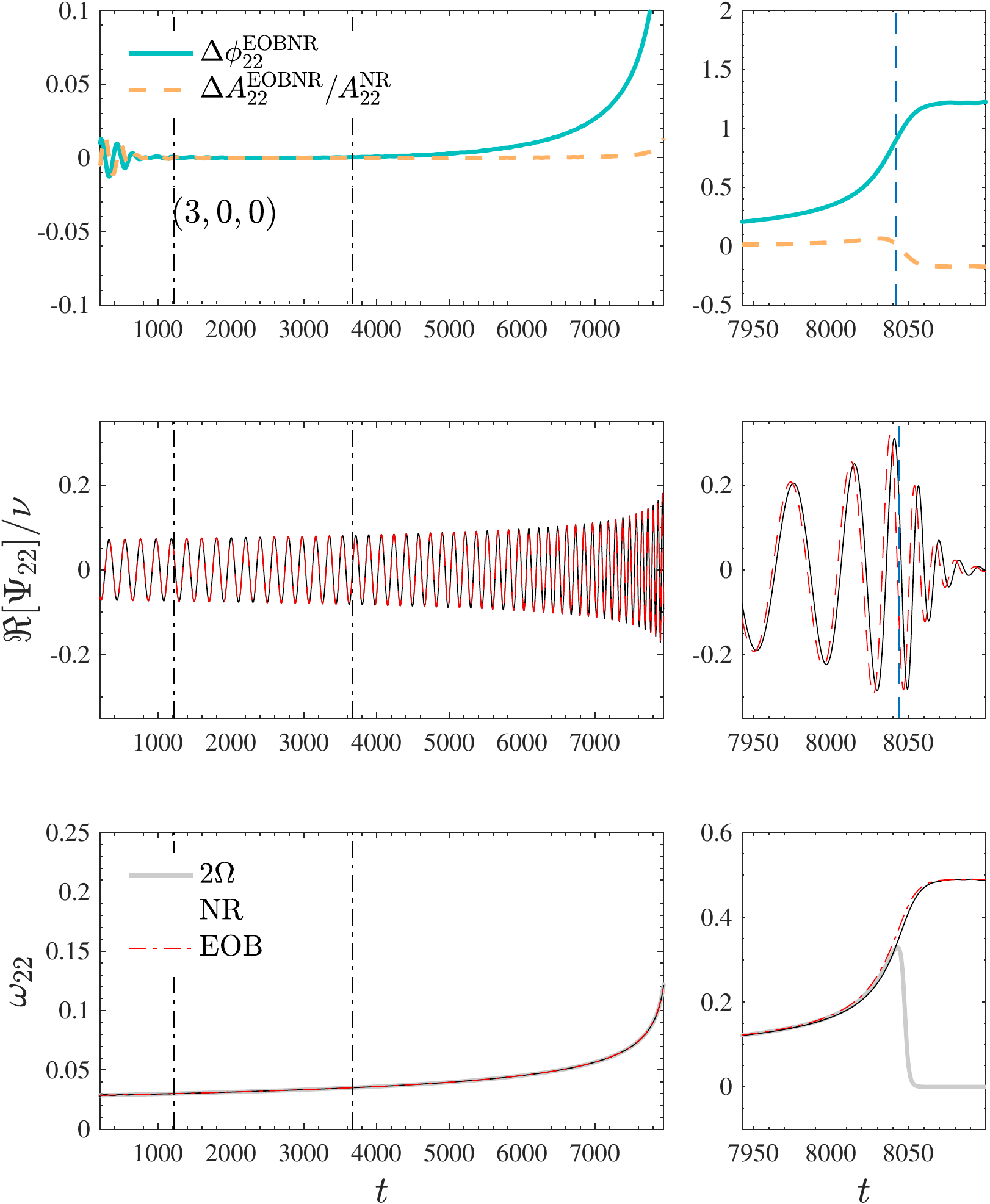}
\caption{\label{fig:q3} Example of acceptable EOB/NR phasing agreement obtained 
suitably selecting the value of $a_6^c$ for $(q,\chi_1,\chi_2)=(3,0,0)$. The NR dataset used
here is SXS:BBH:1221.}
\end{figure}
\begin{figure}[t]
\center
\includegraphics[width=0.45\textwidth]{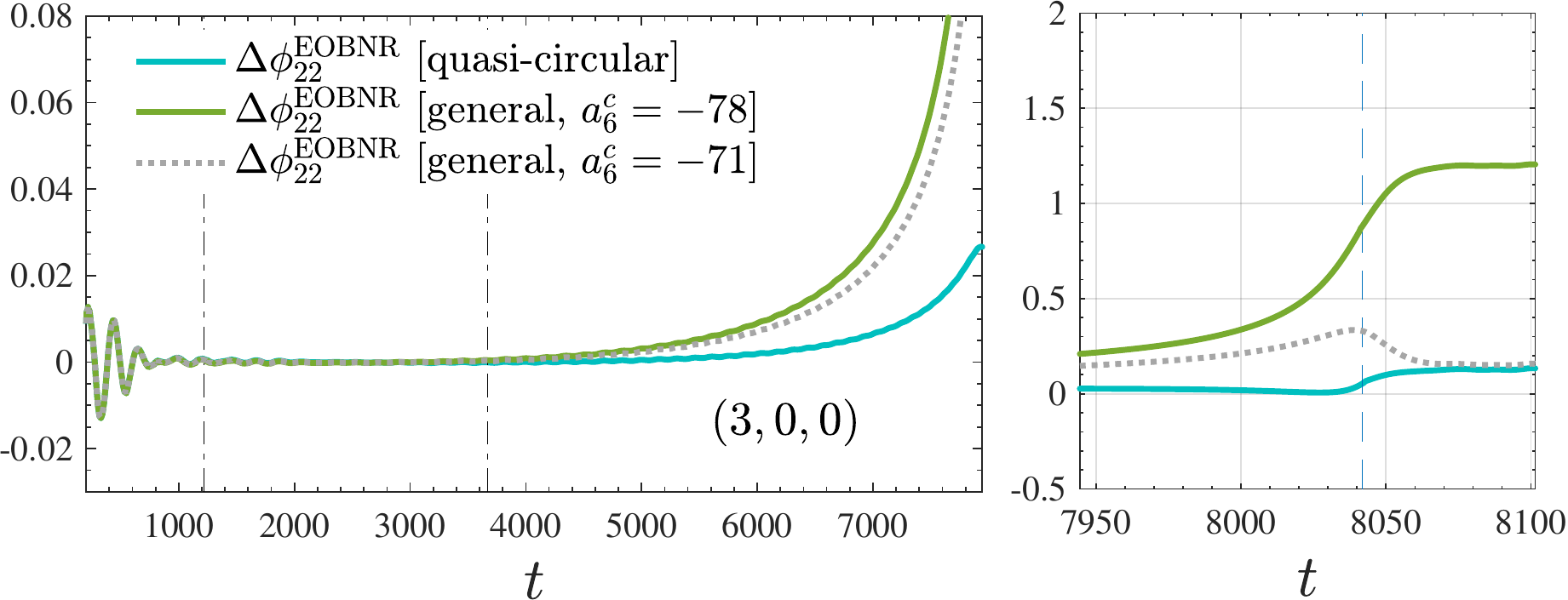}
\caption{\label{fig:q3_Dphi} Comparing phasings for different EOB models or values of $a_6^c$
for $(q,\chi_1,\chi_2)=(3,0,0)$, SXS:BBH:1221. The EOB/NR phase agreement of the general
model (for various values of $a_6^c$, see text) up to merger is acceptable, but is significantly
worse than for the simple quasi-circular model.}
\end{figure}

We now proceed in determining new analytical representations of $(a_6^c,c_3)$.
The procedure is the same as the one discussed in Ref.~\cite{Nagar:2021gss}: 
one first determines the best $a_6^c$ and then the best $c_3$ by analyzing
EOB/NR phasing comparisons. 
Figure~\ref{fig:q3} displays an example of what we consider an acceptable
choice of the parameter, $a_6^c=-78$, informed by the inspection 
of the EOB/NR phase difference $\Delta\phi_{22}^{\rm EOBNR}$. 
When the two waveforms are aligned during the early inspiral 
(the alignment region is indicated by the dashed-dot vertical lines),
the phase difference starts practically flat and then grows monotonically through plunge
and merger until it ends up constant during ringdown. We have 
$\Delta\phi_{22}^{\rm EOBNR}\sim 1$~rad at merger, that is larger than the 
numerical uncertainty ($\sim 0.1$~rad), but it is small enough to yield
values ($< 1\%$) of the EOB/NR unfaithfulness (see below). In addition,
the phase difference accumulated up to merger is still {\it larger} than the one for
the quasi-circular \TEOBResumS{} model, as shown in Fig.~\ref{fig:q3_Dphi}.
However, we explored the flexibility of the model varying $a_6^c$ and we eventually
concluded that within the current analytical setup {\it it is not possible} to further
flatten $\Delta\phi_{22}^{\rm EOBNR}$ and get it close to the quasi-circular case.
The dotted gray line in Fig.~\ref{fig:q3_Dphi} refers to $a_6^c=-71$ and gives an
idea of the flexibility of the model when $a_6^c$ is varied. Increasing $a_6^c$ goes
in the direction of reducing the accumulated phase difference; however, this parameter 
alone is unable to flatten the phase difference at the same level of the 
quasi-circular case (see in particular left panel of Fig.~\ref{fig:q3_Dphi}). 
Moreover, the fact that the phasing is at $~\sim 0.1$ level during the ringdown 
but it is larger up to merger may eventually result in suboptimal values of the 
EOB/NR unfaithfulness. This can also be understood by looking at 
the EOB/NR phase difference when the waves are aligned around merger
and not during the inspiral. Although we could not overcome this problem, we
realized that this is related to structure of $\F_r$, that is given by Eq.~(6) of
Ref.~\cite{Nagar:2021gss}, and in particular to the leading order factor given
by $p_{r_*} u^4$, as obtained  Ref.~\cite{Bini:2012ji}. Some alternative analytical 
expression {\it at leading order} might be worth exploring, although we postpone 
this investigation to future work. In conclusion, when  tuning $a_6^c$ with NR data 
we were careful  to obtain phase differences that are always monotonic versus time, 
in order to reproduce the qualitative structure of the $a_6^c=-78$ phase difference 
in Fig.~\ref{fig:q3_Dphi}. 

\begin{table}[t]
 \caption{\label{tab:a6c} Informing the nonspinning sector of the model. From left to right the columns report: the dataset number, 
 the SXS identification number; the mass ratio $q$; the symmetric mass ratio $\nu$; the first guess value of $a_6^c$ and the fitted
 value from Eq.~\eqref{eq:a6c_fit}.}
   \begin{center}
     \begin{ruledtabular}
\begin{tabular}{c c c c c c} 
$\#$ & SXS & $q$ & $\nu$ &  $a_6^c$  & $a_6^c(\nu)$ \\
  \hline
  \hline
  1 & SXS:BBH:0180  & 1      & 0.25  & $-93$ & $-93.0372$\\  
  2 &  SXS:BBH:1221 & 3  &  0.204 &$-78$ &  $-77.969$\\
  3 &  SXS:BBH:0056 & 6     &  0.139 & $-58$&  $-57.308$\\
  4 & SXS:BBH:0063  & 8     &  0.0988 &$-47$ & $-48.525$ \\ 
  5 & SXS:BBH:0303  & 10   & 0.0826& $-43$ & $-42.161$
  \end{tabular}
 \end{ruledtabular}
 \end{center}
 \end{table}
\begin{figure}[t]
\center
\includegraphics[width=0.45\textwidth]{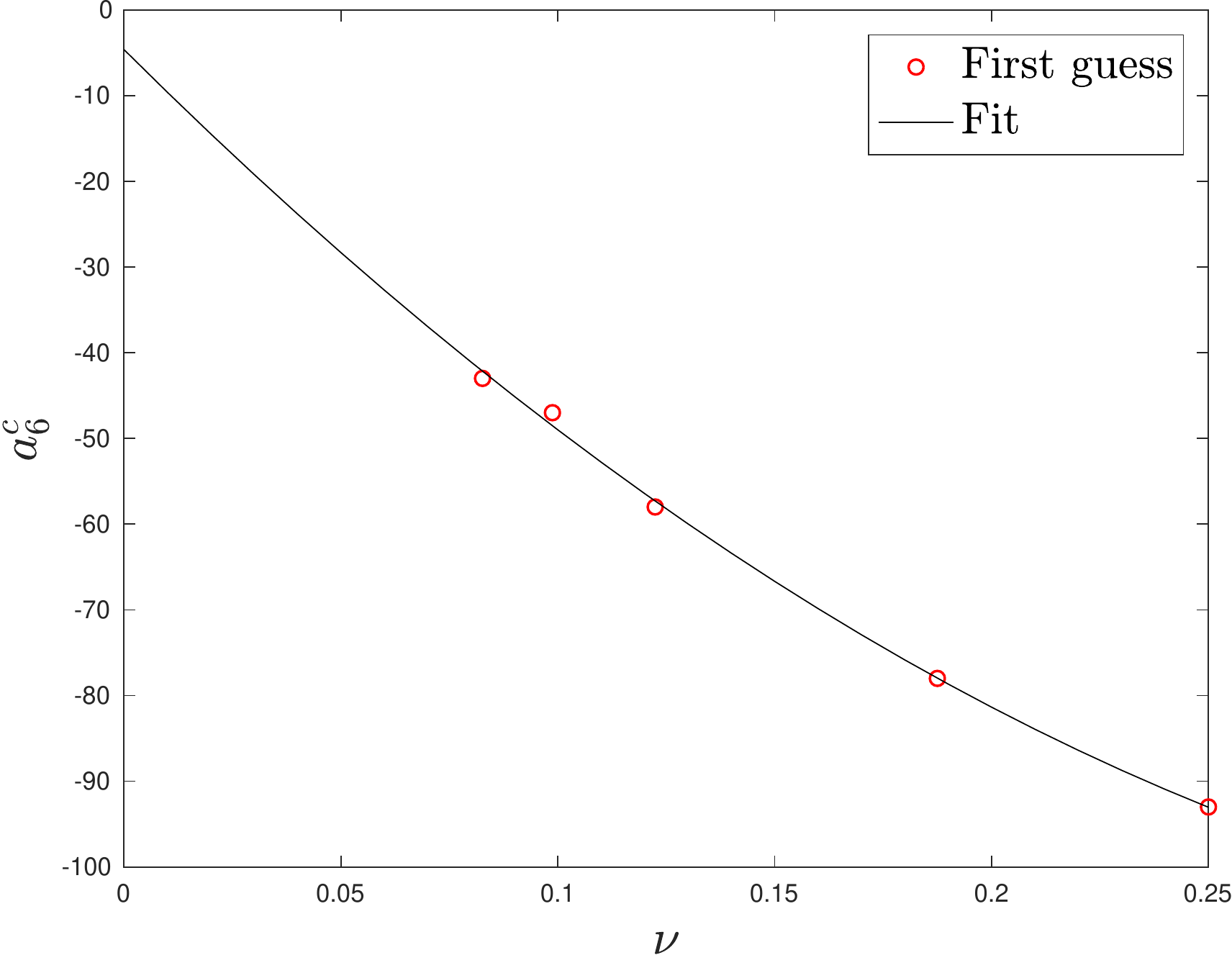}
\caption{\label{fig:a6c} First-guess values of $a_6^c$ from Table~\ref{tab:a6c}. Note the rather simple
functional behavior of $a_6^c(\nu)$ with respect to Ref.~\cite{Nagar:2021gss}, see in particular Eq.~(14) therein.
This fact suggests that the new analytical building blocks have simplified the impact of NR
information in the completion of the model.}
\end{figure}
\begin{table}[t]
   \caption{\label{tab:c3}Informing the spinning sector of the model. From left to right the columns report:
   the dataset number, the SXS identification number; the mass ratio and the individual dimensionless 
   spins $(q,\chi_1,\chi_2)$; the first-guess values of $c_3$ used to inform the global interpolating fit 
   given in Eq.~\eqref{eq:c3fit}, and the corresponding $c_3^{\rm fit}$ values.}
   \begin{center}
 \begin{ruledtabular}
   \begin{tabular}{lllll}
     $\#$ & ID & $(q,\chi_1,\chi_2)$ & $c_3^{\rm first\;guess}$ & $c_3^{\rm fit}$\\
     \hline
    1 & SXS:BBH:0156 &$(1,-0.95,-0.95)$     & 89 & 88.822 \\
    2 & SXS:BBH:0159 &$(1,-0.90,-0.90)$     & 86.5 & 86.538 \\
    3 & SXS:BBH:0154 &$(1,-0.80,-0.80)$     & 81  & 81.508 \\
    4 & SXS:BBH:0215 &$(1,-0.60,-0.60)$     & 70.5 & 70.144 \\
    5 & SXS:BBH:0150 &$(1,+0.20,+0.20)$   & 26.5 & 26.677 \\
    6 & SXS:BBH:0228 &$(1,+0.60,+0.60)$   & 16.0 & 15.765 \\
    7 & SXS:BBH:0230 &$(1,+0.80,+0.80)$   & 13.0 & 12.920 \\
    8 & SXS:BBH:0153 &$(1,+0.85,+0.85)$   & 12.0 & 12.278 \\
    9 & SXS:BBH:0160 &$(1,+0.90,+0.90)$   & 11.5 & 11.595\\
    10 & SXS:BBH:0157 &$(1,+0.95,+0.95)$ & 11.0 & 10.827\\
    11 & SXS:BBH:0004 &$(1,-0.50,0)$         & 54.5 & 46.723 \\
    12 & SXS:BBH:0231 &$(1,+0.90,0)$        & 24.0 &  23.008 \\
    13 & SXS:BBH:0232 &$(1,+0.90,+0.50)$ & 15.8 & 16.082 \\
    14 & SXS:BBH:0005 &$(1,+0.50,0)$        & 34.3 & 27.136 \\
    15 & SXS:BBH:0016 &$(1.5,-0.50,0)$     & 57.0 & 49.654 \\
    16 & SXS:BBH:0016 &$(1.5,+0.95,+0.95)$ & 13.0 & 11.720 \\
    17 & SXS:BBH:0255 &$(2,+0.60,0)$          & 29.0 & 23.147\\
    18 & SXS:BBH:0256 &$(2,+0.60,+0.60)$   & 20.8 & 17.37 \\
    19 & SXS:BBH:0257 &$(2,+0.85,+0.85)$   & 14.7 & 14.56 \\ 
    20 & SXS:BBH:0036 &$(3,-0.50,0)$           &  60.0 & 53.095\\
    21 & SXS:BBH:0267 &$(3,-0.50,-0.50)$     & 69.5 & 60.37\\ 
    22 & SXS:BBH:0174 &$(3,+0.50,0)$           &  30.0 & 24.210\\
    23 & SXS:BBH:0291 &$(3,+0.60,+0.60)$        &  23.4 & 19.635 \\
    24 & SXS:BBH:0293 &$(3,+0.85,+0.85)$        & 16.2 & 17.759 \\
    25 & SXS:BBH:1434 &$(4.368,+0.80,+0.80)$ & 20.3 & 20.715 \\
    26 & SXS:BBH:0060 &$(5,-0.50,0)$                & 62.0 & 55.385\\
    27 & SXS:BBH:0110 &$(5,+0.50,0)$               & 31.0 & 24.488\\
    28 & SXS:BBH:1375 &$(8,-0.90,0)$               & 64.0 & 71.91 \\
    29 & SXS:BBH:0064 &$(8,-0.50,0)$               & 57.0 & 55.385 \\
    30 & SXS:BBH:0065 &$(8,+0.50,0)$              & 28.5 & 24.306\\
 \end{tabular}
 \end{ruledtabular}
 \end{center}
 \end{table}
%
\begin{figure*}[t]
\center
\includegraphics[width=0.4\textwidth]{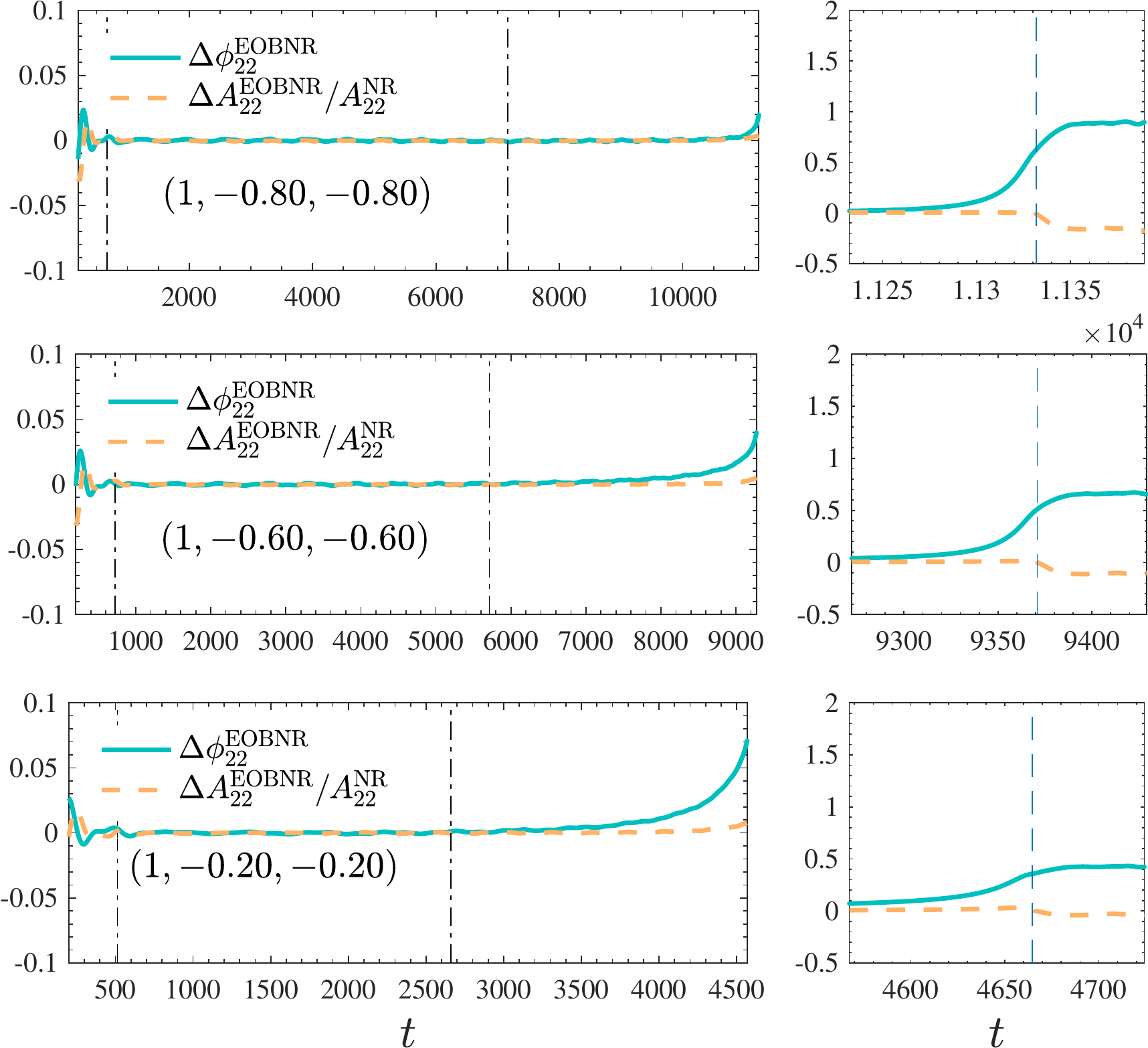}
\qquad
\includegraphics[width=0.4\textwidth]{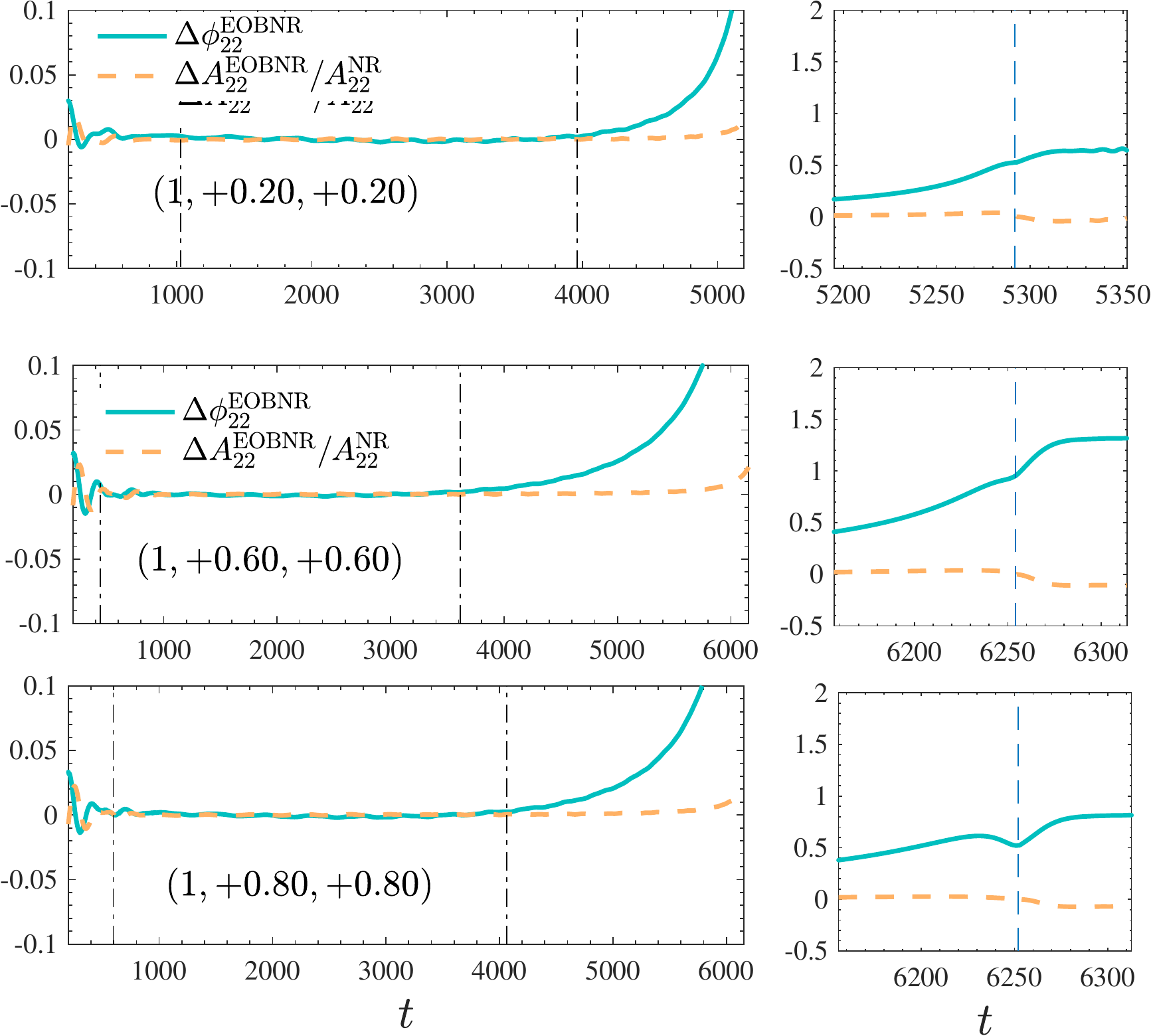}
\caption{\label{fig:q1c3_tuned} EOB/NR time-domain phase difference $\Delta\phi_{22}^{\rm EOBNR}$ and relative amplitude difference,
$\Delta A^{\rm EOBNR}_{22}/A^{\rm NR}_{22}$ for a sample of equal-mass, equal-spin configurations. The model uses here the $c_3$ 
parameter from Eq.~\eqref{eq:c3fit}. The dash-dotted vertical lines indicate the alignment region, while the dashed line indicates the 
merger location. Note that $\bar{F}_{\rm EOBNR}^{\rm max}\lesssim 1\%$ even if  the accumulated phase difference $\sim 1$~rad 
at merger for some configurations.}
\end{figure*}
We determined $a_6^c$ separately for different 5 datasets, listed in 
Table~\ref{tab:a6c}: the fifth column of the table reports our best choices.
When plotted versus $\nu$, one finds a rather simple behavior,
see Fig.~\ref{fig:a6c}, that is fitted as
\begin{equation}
\label{eq:a6c_fit}
a_6^c(\nu) = 599.96\nu^2 - 503.57\nu - 4.6416.
\end{equation}
The fit is very accurate, as shown by the sixth column of Table~\ref{tab:a6c}. In this respect,
it is interesting to note that, although the physics that the model describes is the 
same of the model  of Ref.~\cite{Nagar:2021gss}, the differences in the analytical content and 
in the resummations yield a very simple behavior of $a_6^c(\nu)$. 
This is in striking contrast with  Ref.~\cite{Nagar:2021gss}, where it was needed 
an exponential function to fit (at a lower accuracy level) the single values of $a_6^c$. 

The new functional form of $a_6^c(\nu)$ given by Eq.~\eqref{eq:a6c_fit} calls for a similarly
new determination of the effective spin-orbit parameter $c_3$. We do so using a set of NR data
that is slightly different from the one used in Ref.~\cite{Nagar:2020pcj} so to improve the
robustness of the model in certain corners of the parameter space. The NR datasets used 
are listed in Table~\ref{tab:c3}. Following Ref.~\cite{Nagar:2020pcj}, for each dataset we 
report  the value of $c_3^{\rm first\;guess}$ obtained by comparing
EOB and NR phasing so that the accumulated phase difference is of the order of the NR 
uncertainty (and/or so that consistency between NR and EOB frequencies around merger 
is achieved as much as possible). Similarly to the case of $a_6^c$, the robustness of the model
allows us to efficiently do this by hand without any automatized procedure. 
The $c_3^{\rm first\;guess}$ values of Table~\ref{tab:c3} 
are fitted with a global function of the spin variables $\tilde{a}_i\equiv S_i/(m_i M)$ with $i=1,2$
of the form
\begin{align}
  \label{eq:c3fit}
& c_3(\tilde{a}_1,\tilde{a}_2,\nu)=
p_0\dfrac{1+n_1\tilde{a}_0+n_2\tilde{a}_0^2+n_3\tilde{a}_0^3+n_4\tilde{a}_0^4}{1+d_1\tilde{a}_0}\nonumber\\
&+ p_1 \tilde{a}_0\nu\sqrt{1-4\nu}+ p_2\left(\tilde{a}_1-\tilde{a}_2\right)\nu^2 + p_3 \tilde{a}_0\nu^2\sqrt{1-4\nu},
\end{align}
where $\tilde{a}_0\equiv \tilde{a}_1+\tilde{a}_2$ and the functional form is the same of previous works\footnote{Note that this function is not 
symmetric for exchange of $1\leftrightarrow2$. This can create an ambiguity for $q=1$, so that the value 
of $c_3$ for $(1,0.6,0.4)$ is in fact different from the one for $(1,0.4,0.6)$. In fact, our convention and 
implementations are such that for $q=1$, $\chi_1$ is {\it always} the largest spin.}. 
This term helps in improving the fit flexibility as the mass ratio increases. The fitting coefficients read
\begin{align}
p_0 &=  35.482253,\\ 
 n_1 &=  -1.730483,\\ 
 n_2 &=   1.144438,\\
 n_3 &=   0.098420,\\ 
 n_4 &=  -0.329288,\\ 
 d_1 &=  -0.345207,\\
p_1  &=  244.505,\\
p_2 &= 148.184,\\
p_3  &= -1085.35.
\end{align}

\begin{figure}[t]
\center
\includegraphics[width=0.5\textwidth]{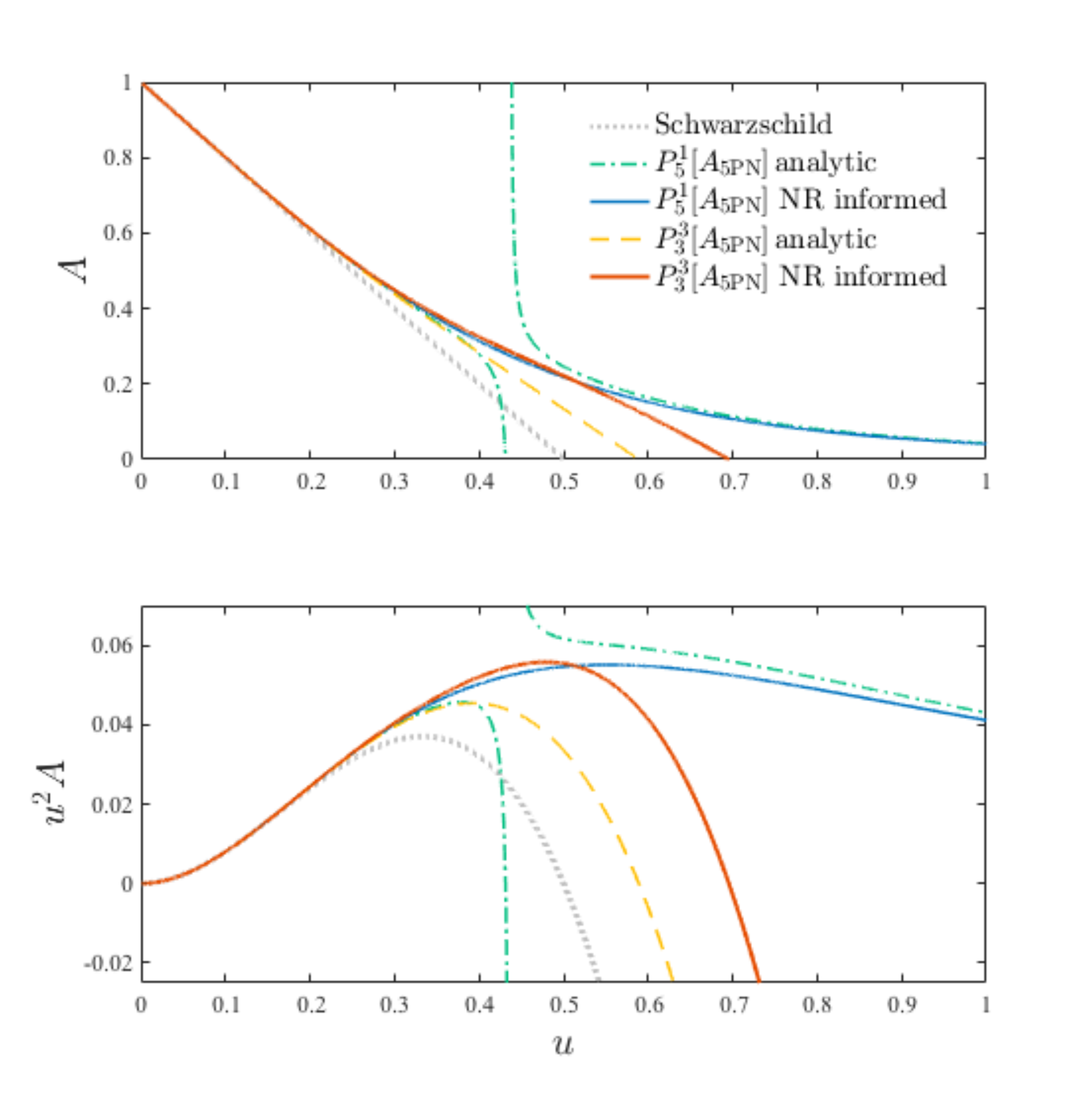}
\caption{\label{fig:Au2A} Comparison between different $A$ functions (top panel) and effective photon potentials, $u^2 A$
(bottom panel) for $q=1$. The picture highlights that the commonly used $P^1_5[A]$ approximant develops an unphysical
pole when $a_6^c=a_{6 \rm anlyt}^c$ from Eq.~\eqref{eq:a6c_anlyt}. One also notices the consistency between the two
NR-informed $A$ functions up to the effective light-ring, i.e. the peak of the function $u^2 A$. 
Note however that the difference between the potentials around $u\simeq 0.4$ is $\sim 10^{-2}$, so that they 
do actually yield different EOB dynamics.}
\end{figure}

%
\begin{figure}[t]
\center
\includegraphics[width=0.44\textwidth]{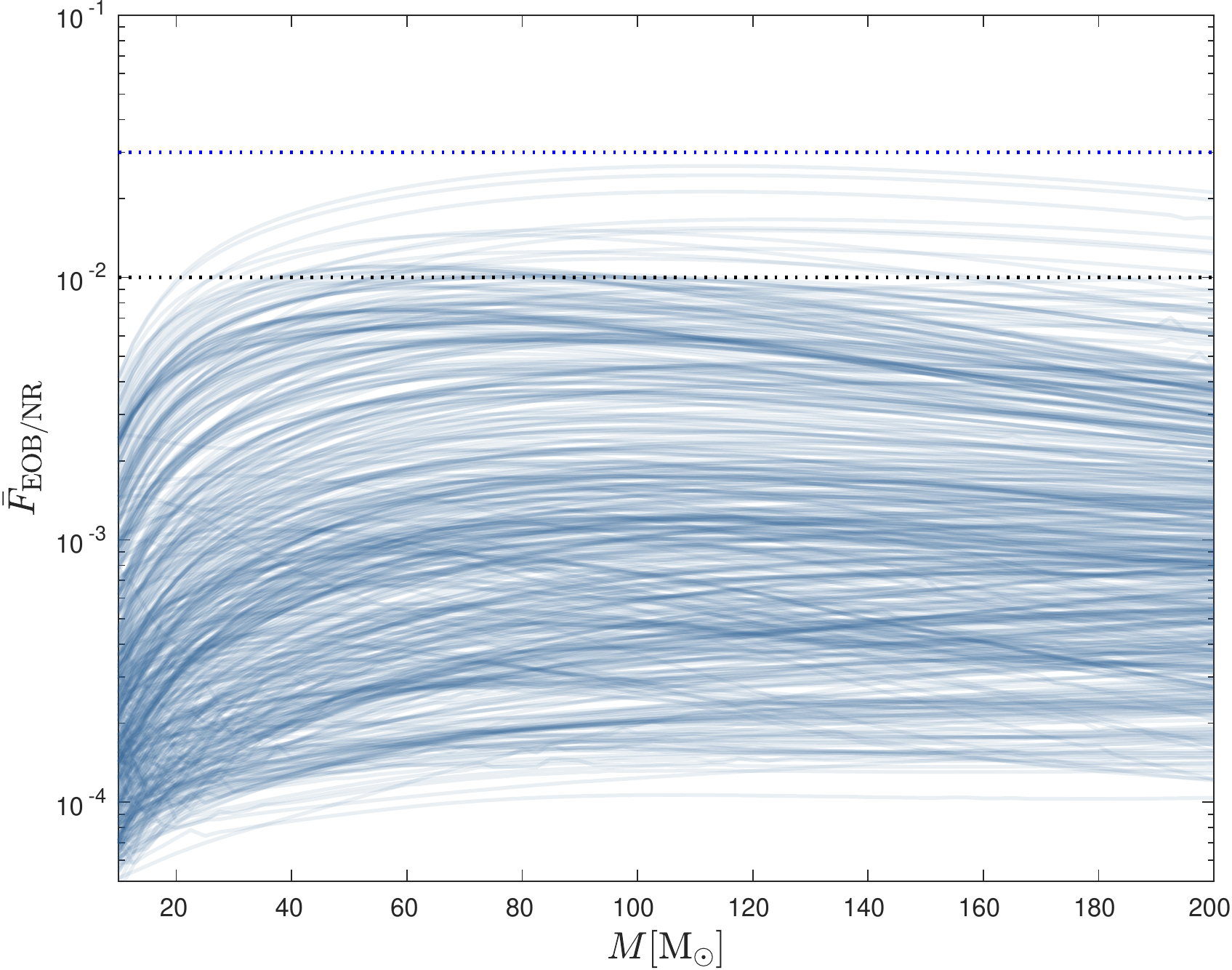}
\caption{\label{fig:barF_circular} Unfaithfulness $\bar{F}_{\rm EOB/NR}(M)$ between the quasi-circular limit of the general 
           \TEOBResumS{} model and the complete SXS catalog of non-eccentric and non-precessing (spin-aligned) waveforms. 
            These results are obtained using the NR-tuned N$^3$LO spin-orbit contribution. The horizontal dotted lines mark the $0.01$ (black) 
            and $0.03$ (blue) values.}
\end{figure}

\begin{figure}[t]
\center
\includegraphics[width=0.44\textwidth]{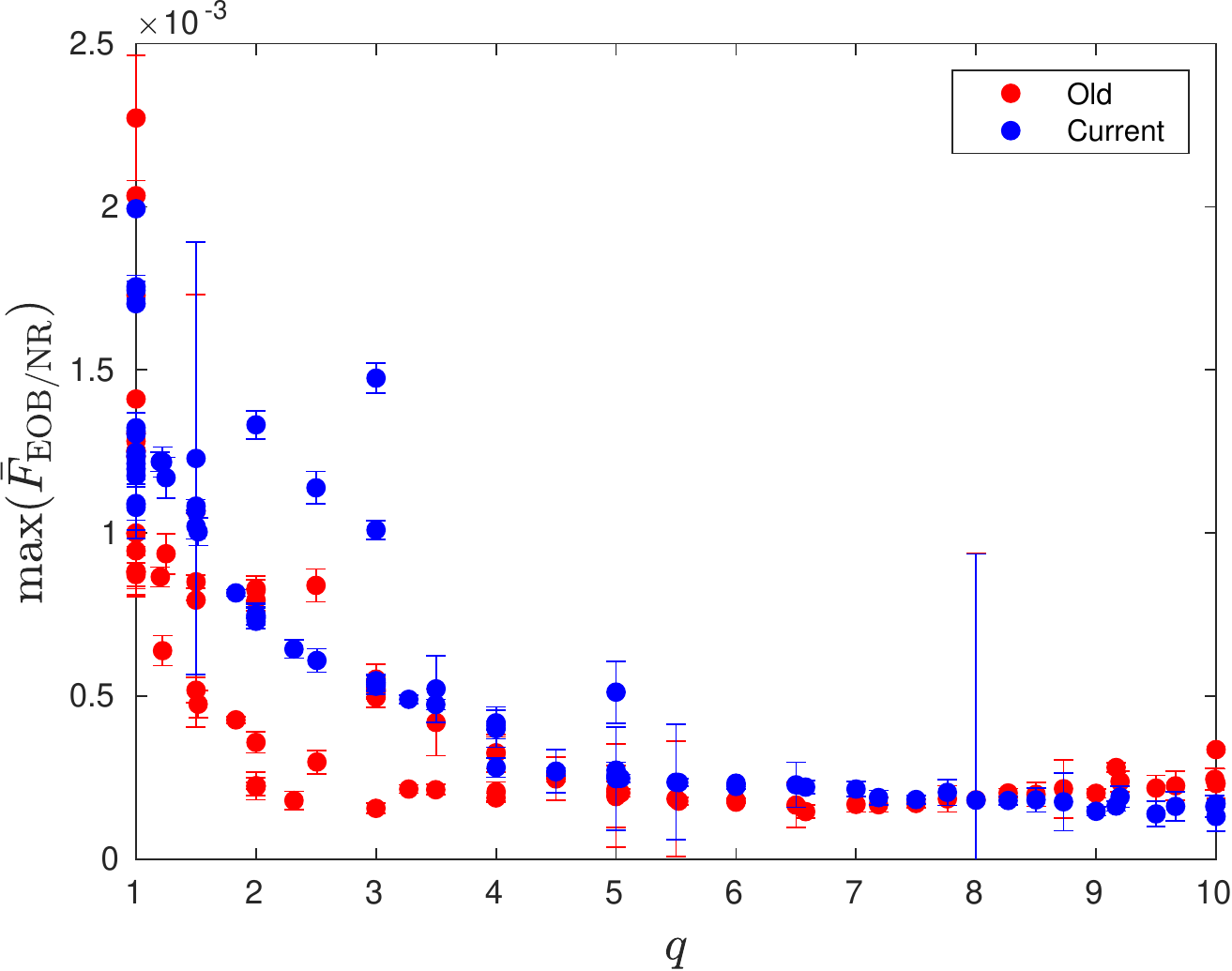}
\caption{\label{fig:maxF_s0}
Complement to Fig.~\ref{fig:barF_circular}: 
$\bar{F}^{\rm max}_{\rm EOB/NR}$ values for all SXS nonspinning configurations.
The current iteration of \TEOBResumS{} is compared to the quasi-circular model of Ref.~\cite{Nagar:2020pcj}, 
with errorbars that represent the values of $\bar{F}^{\rm max}_{\rm NR/NR}$ cited therein
(see Tables XVIII-XIX).
Note that SXS datasets with the same $q$ have different NR accuracies.
The EOB/NR agreement for this subset of data is largely improved with respect to Fig.~3 of Ref.~\cite{Nagar:2021gss}
and mostly consistent with the result of the quasi-circular model.}
\end{figure}

\subsubsection{Consistency between EOB potentials}
Now that we have determined the new expression of $a_6^c(\nu)$ it is instructive to compare different
realization of the potential. The top panel of Fig.~\ref{fig:Au2A} shows together different curves for the
case $q=1$:
(i) The $P^3_3[A]$ potential with the NR-informed $a_6^c(\nu)$ given by Eq.~\eqref{eq:a6c_fit} above;
(ii) the $P^1_5[A]$ potential of \TEOBResumS{}, where the NR-informed function is given by Eq.~(33) of
Ref.~\cite{Nagar:2020pcj}; (iii) the $P^1_5[A]$ function with $a_6^c=a^c_{6\rm anlyt}$; 
(iv) the $P^3_3[A]$ function  with $a_6^c=a^c_{6\rm anlyt}$. In the bottom panel of the figure we
show the effective photon potential $u^2 A$. 
The most interesting outcome is the visual {\it consistency} between 
the two NR-informed potentials up to $u\simeq 0.4$.
This reflects in two close dynamics, that eventually yield highly faithful EOB/NR phasing for the
nonspinning case, as we will see below. The fact is remarkable because
{\it both} the radiation reaction {and} the $(D,Q)$ potentials are different in
the two cases.
One should note, however, that the fact that the two potentials are consistent
up to $u=0.4$ {\it does not mean} that they are equivalent and that the conservative
dynamics coincide. In fact, it is known~\cite{Damour:2012ky} that two $A$ potentials
are equivalent when their difference is of the order of $10^{-4}$. 
The two NR-informed potentials differ by just $10^{-2}$, so that even if they
look close, they are meaningfully different.
A similar visual consistency shows up also for the  $P^1_5[A]$ analytical function, 
despite the presence  of the spurious pole. By contrast the fully 
analytical $P^3_3[A]$ is significantly separated from the others.
In practical terms, when used in the EOB dynamics, the $P^3_3[A]$ analytical 
potential will {\it accelerate}  the inspiral with respect to the NR-informed ones, 
eventually yielding unacceptably large phase differences at merger. 
If one wished to incorporate this specific resummation, some other element of the model
(e.g. radiation reaction or the $(D,Q)$ functions) should be modified to balance its attractive effect.
This gives a pedagogical example of the fact that the accessibility of high-order
PN information\footnote{Although incomplete, seen the lack of the yet 
unknown $(a_6^{\nu^2},d_5^{\nu^2})$ coefficients.} does not necessarily 
simplify or help the construction of waveform models and it is pragmatically
more efficient to resort to NR-informed functions.

\begin{figure*}[t]
\center
\includegraphics[width=0.4\textwidth]{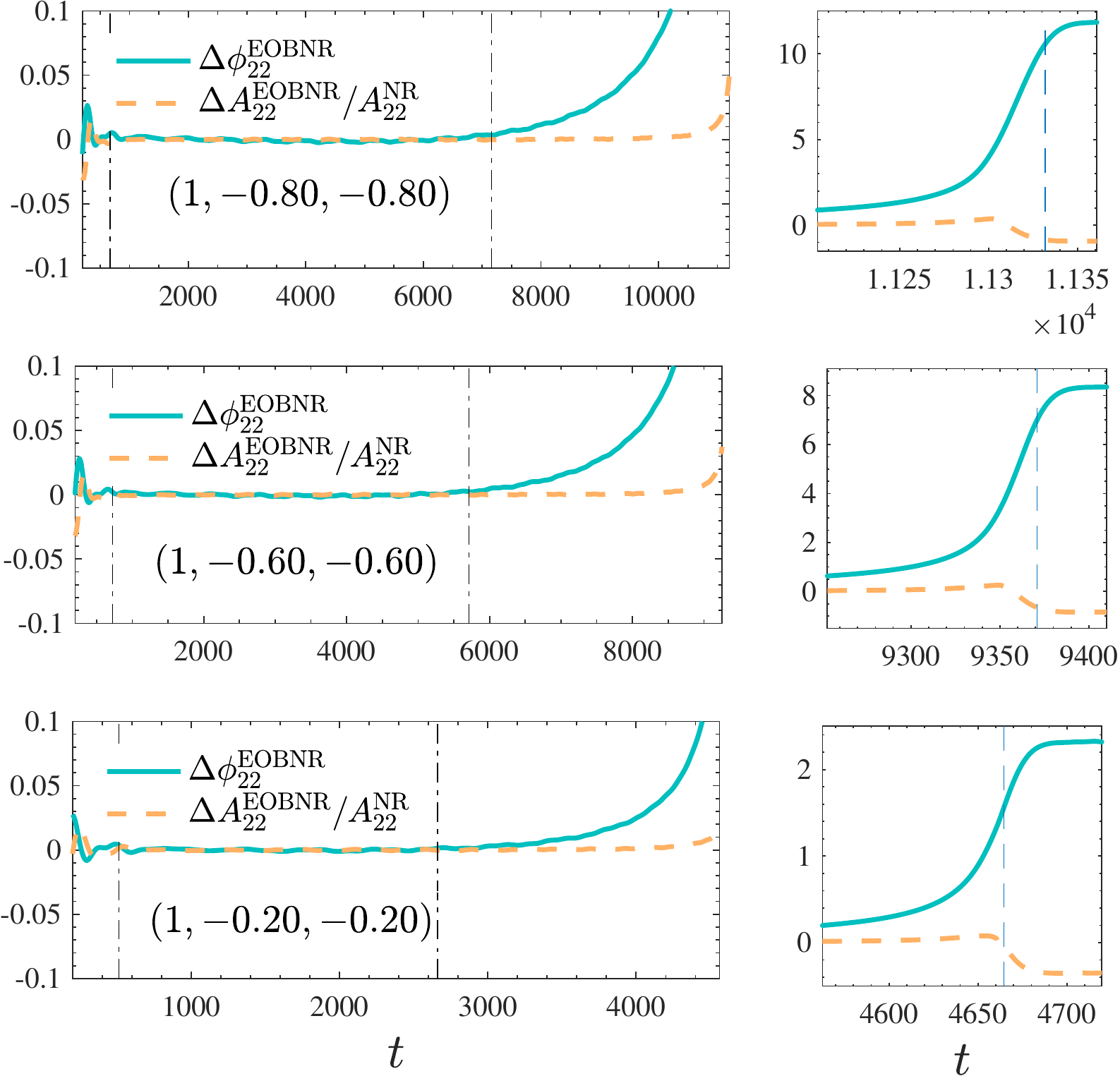}
\qquad
\includegraphics[width=0.4\textwidth]{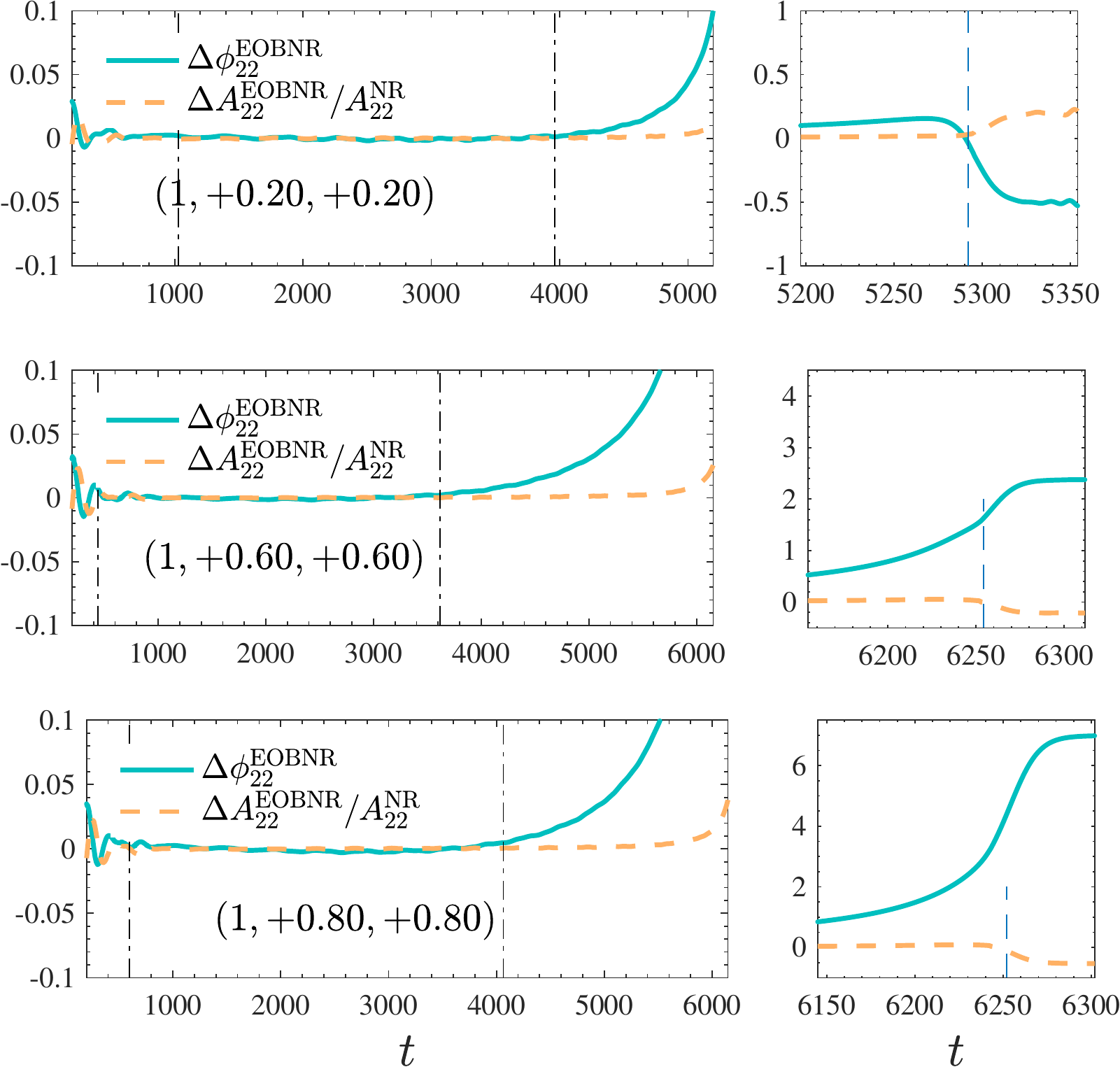}
\caption{\label{fig:q1_c40}EOB/NR time-domain phase difference $\Delta\phi_{22}^{\rm EOBNR}$ and relative amplitude difference,
$\Delta A^{\rm EOBNR}_{22}/A^{\rm NR}_{22}$ for a sample of equal-mass, equal-spin configurations. The model uses here the 
{\it analytical description} of N$^3$LO spin-orbit effect, with $c_4=0$. The dash-dotted vertical lines indicate the alignment region, 
while the dashed line indicates the merger location. The large values of $\Delta\phi_{22}^{\rm EOBNR}$ at merger eventually
end up with values of  $\bar{F}_{\rm EOBNR}^{\rm max}$ even above the $3\%$ level, see Fig.~\ref{fig:barF_circ_cN3LO} below.}
\end{figure*}

%
\begin{figure}[t]
\center
\includegraphics[width=0.44\textwidth]{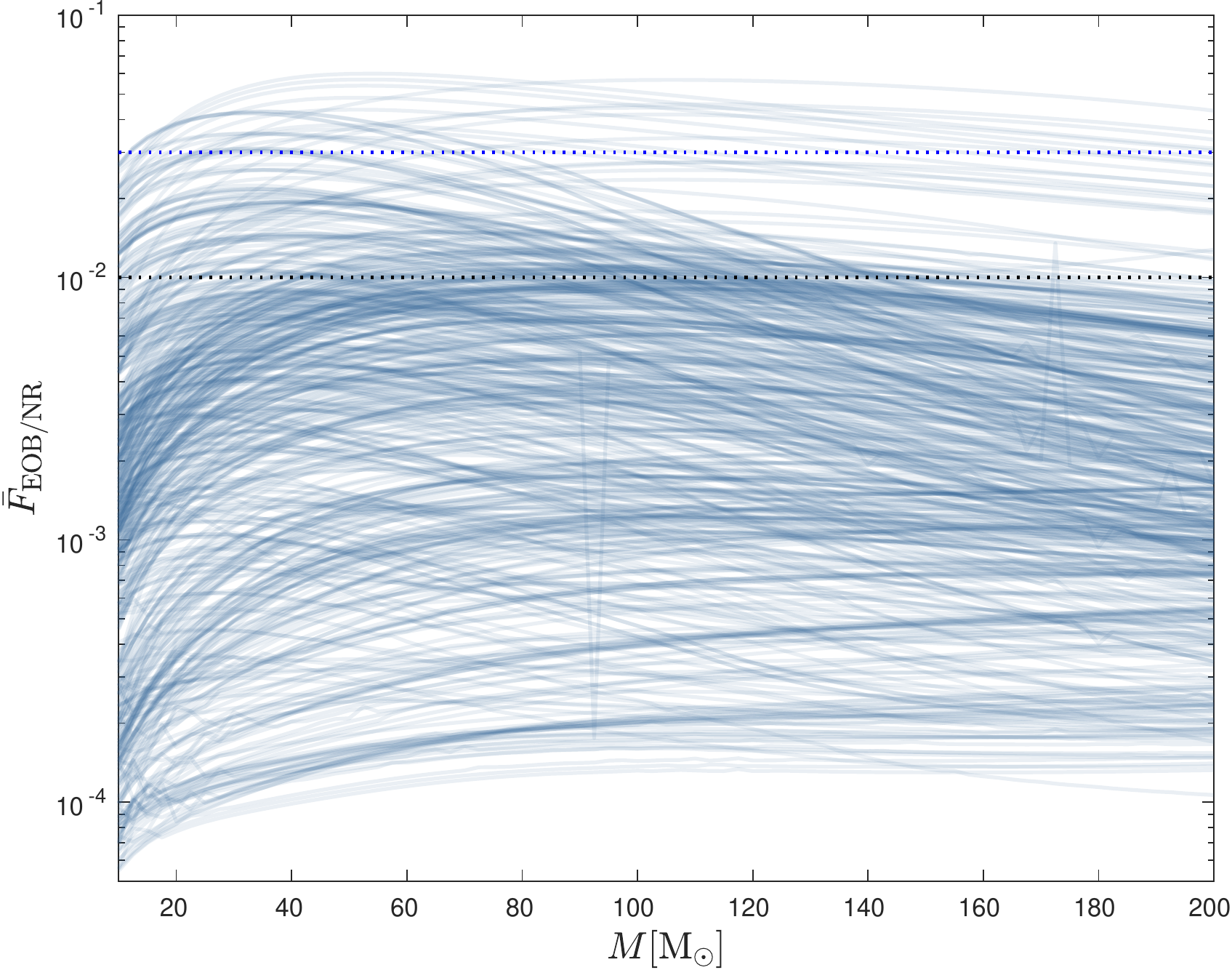}
\caption{\label{fig:barF_circ_cN3LO} Unfaithfulness $\bar{F}_{\rm EOB/NR}(M)$ between \TEOBResumS{} and the 
            complete SXS catalog of non-eccentric non-precessing waveforms obtained using the analytical N$^3$LO 
            spin-orbit contribution. The horizontal dotted lines mark the $0.01$ (black) and $0.03$ (blue) values.}
\end{figure}
%
\begin{figure}[t]
\center
\includegraphics[width=0.44\textwidth]{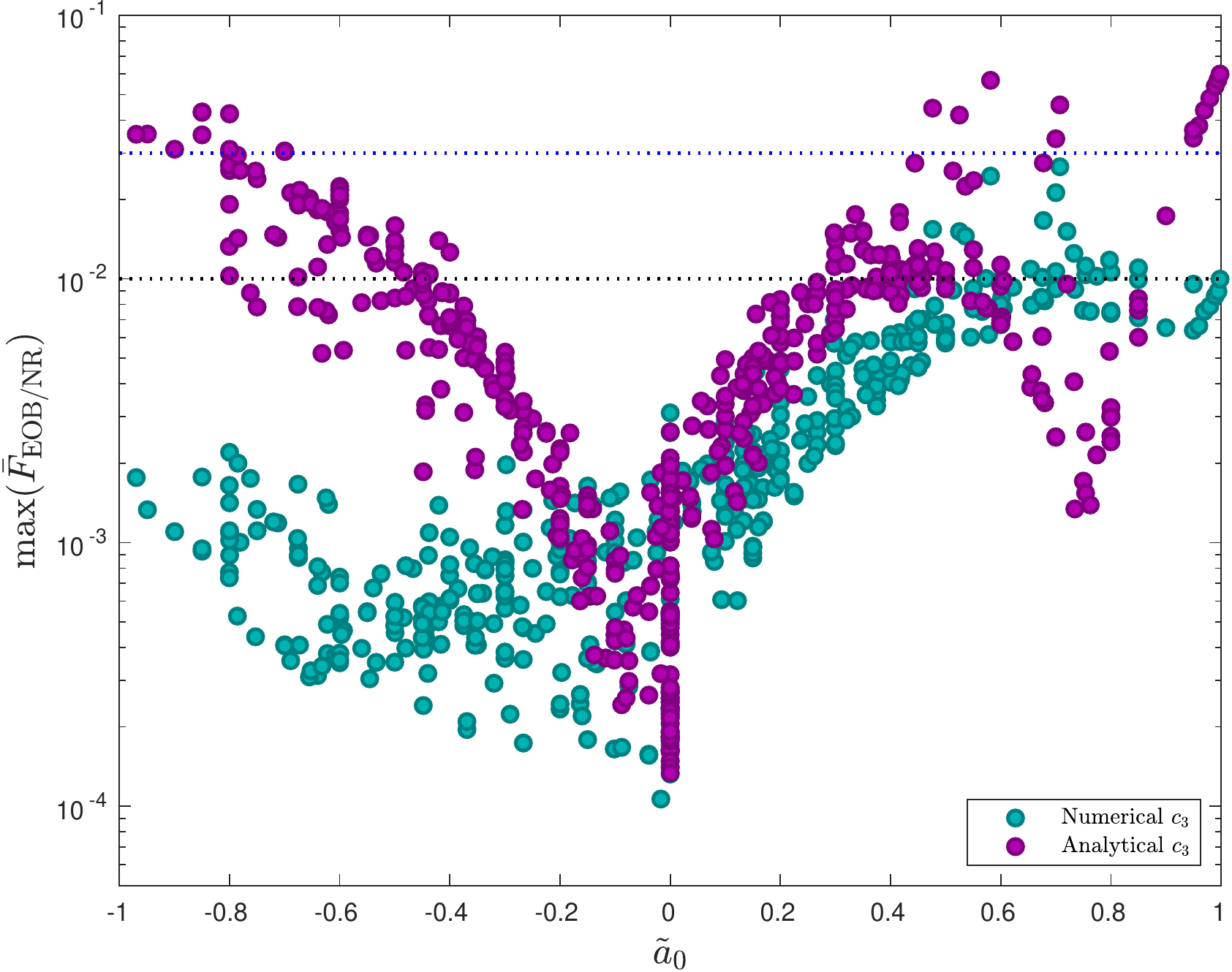} 
\caption{\label{fig:max_barF} Global picture of the maximum EOB/NR unfaithfulness 
from Fig.~\ref{fig:barF_circular} and Fig.~\ref{fig:barF_circ_cN3LO} using the NR-informed and the analytical N$^3$LO 
spin-orbit contribution respectively. The black and blue dotted lines mark the $0.01$ and $0.03$ 
values respectively. The use of the analytical spin-orbit contribution delivers a 
NR-faithful model only in a rather limited range of $\tilde{a}_0$.}
\end{figure}

%
\begin{figure*}[t]
\center
\includegraphics[width=0.4\textwidth]{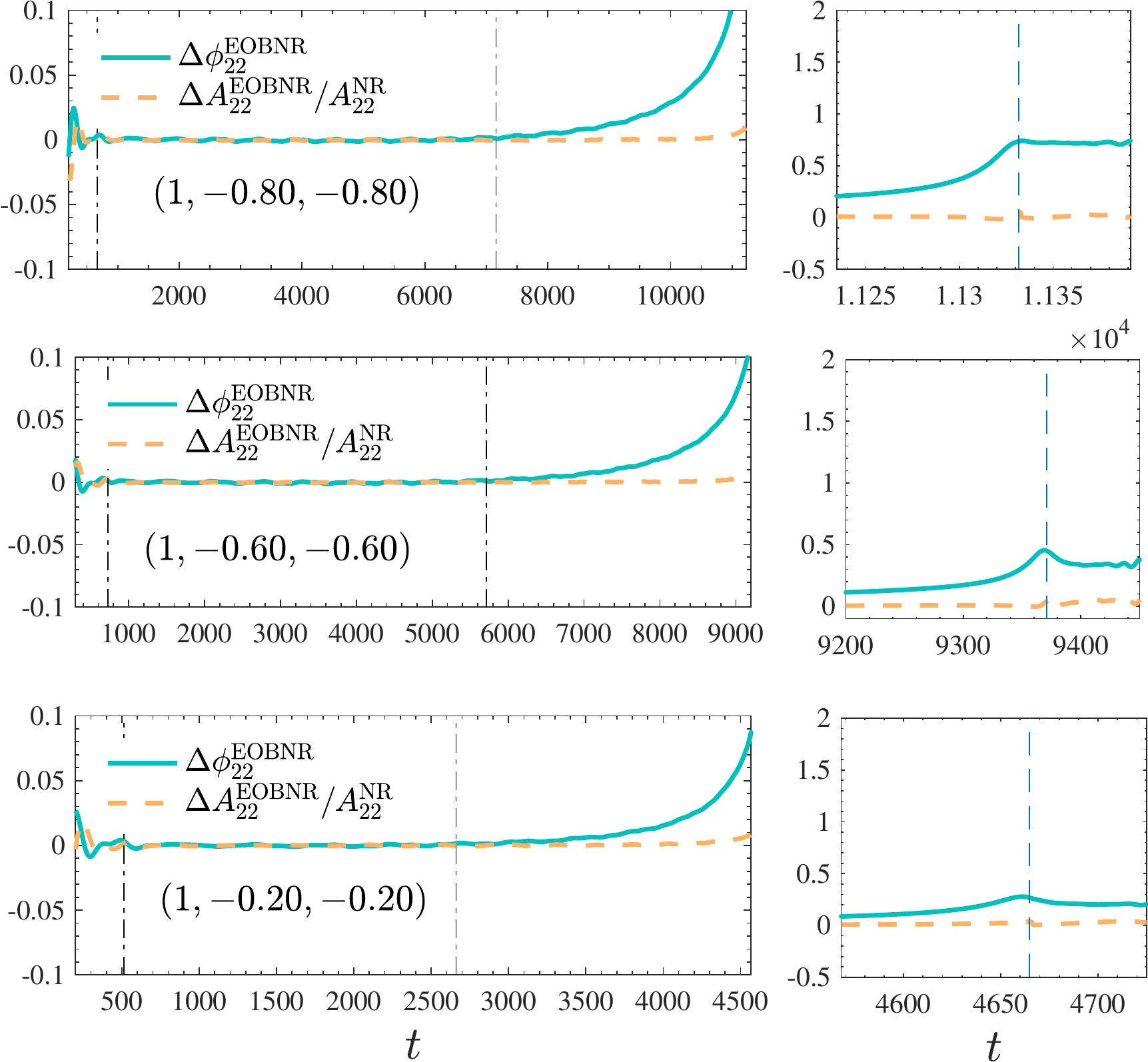}
\qquad
\includegraphics[width=0.4\textwidth]{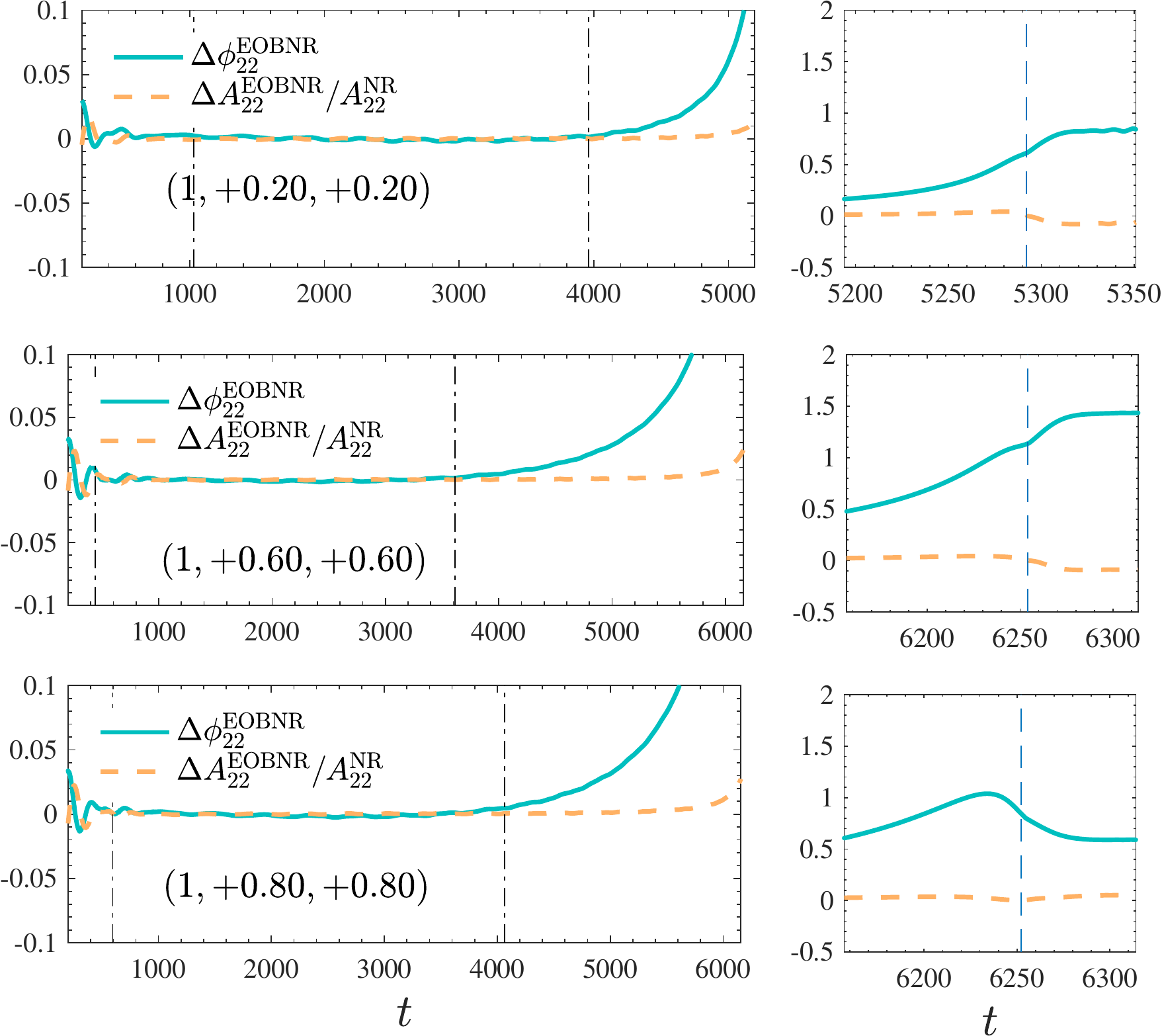}
\caption{\label{fig:q1c4_tuned} EOB/NR time-domain phase difference $\Delta\phi_{22}^{\rm EOBNR}$ and relative amplitude difference,
$\Delta A^{\rm EOBNR}_{22}/A^{\rm NR}_{22}$ for a sample of equal-mass, equal-spin configurations. The model uses here the 
{\it analytical description} of N$^3$LO spin-orbit effect augmented by a NR-tuned coefficient $c_4$ at N$^4$LO, 
as given by Eq.~\eqref{eq:c4}. The dash-dotted vertical lines indicate the alignment region, while the dashed line indicates 
the merger location. The phase difference at merger is reduced with respect to the $c_4=0$ case of Fig.~\ref{fig:q1_c40}, 
although it is still slightly less good (especially for negative spins) than the simple NR-tuned $c_3$ case of Fig.~\ref{fig:q1c3_tuned}.}
\end{figure*}

\subsubsection{Validating the model}
To evaluate the quality of the EOB waveform we computed the EOB/NR unfaithfulness
 weighted by the Advanced LIGO noise over all available spin-aligned SXS configurations. 
 Considering two waveforms $(h_1,h_2)$, the unfaithfulness is a function of the total mass $M$ 
of the binary and is defined as
\be
\label{eq:barF}
\bar{F}(M) \equiv 1-F=1 -\max_{t_0,\phi_0}\dfrac{\langle h_1,h_2\rangle}{||h_1||||h_2||},
\ee
where $(t_0,\phi_0)$ are the initial time and phase. We used $||h||\equiv \sqrt{\langle h,h\rangle}$,
and the inner product between two waveforms is defined as 
$\langle h_1,h_2\rangle\equiv 4\Re \int_{f_{\rm min}^{\rm NR}(M)}^\infty \tilde{h}_1(f)\tilde{h}_2^*(f)/S_n(f)\, df$,
where $\tilde{h}(f)$ denotes the Fourier transform of $h(t)$, $S_n(f)$ is the zero-detuned,
high-power noise spectral density of Advanced LIGO~\cite{aLIGODesign_PSD} and
$f_{\rm min}^{\rm NR}(M)=\hat{f}^{\rm NR}_{\rm min}/M$ is the initial frequency of the
NR waveform at highest resolution, i.e. the frequency measured after the junk-radiation
initial transient. Waveforms are tapered in the time-domain so as to reduce high-frequency 
oscillations in the corresponding Fourier transforms. The EOB/NR unfaithfulness is addressed as 
$\bar{F}_{\rm EOB/NR}$. 
The result of this computation is shown in Fig.~\ref{fig:barF_circular}.
We can see that the maximum unfaithfulness is mostly below $0.01$ and always below $0.03$.
We will highlight in Fig.~\ref{fig:max_barF} below that the higher $\bar{F}_{\rm EOB/NR}$ 
correspond to configuration with large spin values, aligned with the orbital angular momentum.

Before doing so, it is instructive also to show separately the $\bar{F}_{\rm EOB/NR}^{\rm max}$
restricted to the nonspinning case. The chosen NR waveforms are listed in 
Tables XVIII-XIX of Ref.~\cite{Nagar:2020pcj}, with the exclusion of the 3 BAM~\cite{Husa:2015iqa} 
ones and 6 precessing configurations that were erroneously included 
there \footnote{Namely SXS:BBH:0850, SXS:BBH:0858, SXS:BBH:0869, SXS:BBH:2019, SXS:BBH:2025 
and SXS:BBH:2030.}.
To better appreciate the improvement with respect to Ref.~\cite{Nagar:2021gss},
Fig.~\ref{fig:maxF_s0} compares the current (nonspinning) $\bar{F}_{\rm EOB/NR}^{\rm max}$ 
values with those of \TEOBResumS{} obtained in Ref.~\cite{Nagar:2020pcj}. There is an excellent 
consistency between the two dataset, although the current model is performing slightly
less well up to $q=4$. This is expected in view of the discussion around Fig.~\ref{fig:q3_Dphi}.

\subsection{Fully analytical effective one body spin-orbit dynamics at N$^3$LO and beyond}
\label{sec:cN3LO}
\begin{table}[t]
   \caption{\label{tab:c4}Informing the N$^4$LO spin-orbit effective contribution. From left to right the columns report:
   the dataset number, the SXS identification number; the mass ratio and the individual dimensionless 
   spins $(q,\chi_1,\chi_2)$; the first-guess values of $c_4$ used to inform the global interpolating fit 
   given in Eq.~\eqref{eq:c3fit}, and the corresponding $c_4^{\rm fit}$ values.}
   \begin{center}
 \begin{ruledtabular}
   \begin{tabular}{llllcc}
     $\#$ & ID & $(q,\chi_1,\chi_2)$& $\tilde{a}_0$ & $c_4^{\rm first\;guess}$ & $c_4^{\rm fit}$ \\
     \hline
    1 & SXS:BBH:0156 &$(1,-0.95,-0.95)$     & $-0.95$ & 230 & 235.99 \\
    2 & SXS:BBH:0154 &$(1,-0.80,-0.80)$     & $-0.80$ &210  &  204.38\\
    3 & SXS:BBH:0215 &$(1,-0.60,-0.60)$     & $-0.60$ &163 & 164.98 \\
    4 & SXS:BBH:         &$(1,-0.20,-0.20)$     & $-0.20$ &107.5 & 95.66 \\
    4 & SXS:BBH:0150 &$(1,+0.20,+0.20)$  & $+0.20$ & 28      & 38.95 \\
    5 & SXS:BBH:0228 &$(1,+0.60,+0.60)$  & $+0.60$ & $-8$   &  $-5.14$ \\
    6 & SXS:BBH:0230 &$(1,+0.80,+0.80)$  & $+0.80$ & $-21$  &  $-22.45$ \\
    10 & SXS:BBH:0157 &$(1,+0.95,+0.95)$ &$+0.95$ & $-30.5$ & $-33.36$ \\
 \end{tabular}
 \end{ruledtabular}
 \end{center}
 \end{table}
\begin{table*}[t]
   \caption{\label{tab:SXS} SXS simulations with eccentricity analyzed in this work. From left to right: the 
   ID of the simulation; the mass ratio $q\equiv m_1/m_2\geq 1$ and the individual dimensionless 
   spins $(\chi_1,\chi_2)$; the time-domain NR phasing uncertainty at merger $\delta\phi^{\rm NR}_{\rm mrg}$; 
   the estimated NR eccentricity at first apastron $e_{\omega_a}^{\rm NR}$; 
   the NR frequency of first apastron $\omega_{a}^{\rm NR}$; 
   the initial EOB eccentricity $e^{\rm EOB}_{\omega_a}$ and apastron frequency $\omega_{a}^{\rm EOB}$ used to start the EOB evolution; 
   the maximal NR unfaithfulness uncertainty, $\bar{F}^{\rm max}_{\rm NR/NR}$ and the maximal EOB/NR unfaithfulness, 
   $\bar{F}_{\rm EOB/NR}^{\rm max}$. }
   \begin{center}
     \begin{ruledtabular}
\begin{tabular}{c| c  c c c c |l l| c c} 
  $\#$ & id & $(q,\chi_1,\chi_2)$ & $\delta\phi^{\rm NR}_{\rm mrg}$[rad]& $e^{\rm NR}_{\omega_a}$ & $\omega_a^{\rm NR}$ &$e^{\rm EOB}_{\omega_a}$ & $\omega_{a}^{\rm EOB}$ & $\bar{F}_{\rm NR/NR}^{\rm max}[\%]$  &$\bar{F}_{\rm EOB/NR}^{\rm max}[\%]$ \\
  \hline
  \hline
1 & SXS:BBH:1355 & $(1,0, 0)$ & $+0.92$ & 0.0620  & 0.03278728 & 0.0888    & 0.02805750  & 0.012 & 0.13 \\
2 & SXS:BBH:1356 & $(1,0, 0)$& $+0.95$ & 0.1000 &  0.02482006  & 0.15038  & 0.019077  & 0.0077 & 0.17\\
3 & SXS:BBH:1358 & $(1,0, 0)$& $+0.25$ & 0.1023 & 0.03108936 & 0.18078   & 0.021238 & 0.016 & 0.17 \\
4 & SXS:BBH:1359 & $(1,0, 0)$& $+0.25$  & 0.1125 & 0.03708305 & 0.18240   & 0.02139 & 0.0024& 0.13 \\
5 & SXS:BBH:1357 & $(1,0, 0)$& $-0.44$  & 0.1096 & 0.03990101 & 0.19201   & 0.01960 & 0.028& 0.11 \\
6 & SXS:BBH:1361 & $(1,0, 0)$& +0.39    & 0.1634 & 0.03269520  & 0.23557   & 0.0210   & 0.057& 0.38\\
7 & SXS:BBH:1360 & $(1,0, 0)$& $-0.22$ & 0.1604 & 0.03138220 & 0.2440  & 0.01953   &0.0094  & 0.31 \\
8 & SXS:BBH:1362 & $(1,0, 0)$& $-0.09$ & 0.1999 & 0.05624375 & 0.3019     & 0.01914 & 0.0098 & 0.13 \\
9 & SXS:BBH:1363 & $(1,0, 0)$& $+0.58$ & 0.2048 & 0.05778104 &  0.30479    & 0.01908 & 0.07 & 0.25 \\
10 & SXS:BBH:1364 & $(2,0, 0)$& $-0.91$ & 0.0518 &  0.03265995   & 0.08464    & 0.025231   & 0.049  & 0.12\\
11 & SXS:BBH:1365 & $(2,0, 0)$& $-0.90$ & 0.0650  &  0.03305974   & 0.11015     & 0.023987 & 0.027& 0.096 \\
12 & SXS:BBH:1366 & $(2,0, 0)$& $-6\times 10^{-4}$ & 0.1109 & 0.03089493 & 0.1496   & 0.02580 &  0.017  & 0.30 \\
13 & SXS:BBH:1367 & $(2,0, 0)$& $+0.60$ & 0.1102 & 0.02975257 & 0.15065    & 0.026025  & 0.0076  & 0.18 \\
14 & SXS:BBH:1368 & $(2,0, 0)$& $-0.71$ & 0.1043 & 0.02930360 &  0.14951  & 0.02527    & 0.026 & 0.34\\
15 & SXS:BBH:1369 & $(2,0, 0)$& $-0.06$ & 0.2053 & 0.04263738 & 0.3134     & 0.01735  & 0.011& 0.21 \\
16 & SXS:BBH:1370 & $(2,0, 0)$& $+0.12$ & 0.1854 &  0.02422231 &  0.3149  & 0.01688  & 0.07& 0.46 \\
17 & SXS:BBH:1371 & $(3,0, 0)$& $+0.92$ & 0.0628 & 0.03263026  & 0.0912     & 0.029058   & 0.12  & 0.17 \\
18 & SXS:BBH:1372 & $(3,0, 0)$& $+0.01$& 0.1035 & 0.03273944 & 0.14915      & 0.026070 & 0.06  & 0.07\\
19 & SXS:BBH:1373 & $(3,0, 0)$& $-0.41$ & 0.1028 & 0.03666911 & 0.15035    & 0.0253 & 0.0034 & 0.17 \\
20 & SXS:BBH:1374 & $(3,0, 0)$& $+0.98$ & 0.1956  & 0.02702594 & 0.314   & 0.016938   & 0.067 & 0.08 \\
\hline
21 & SXS:BBH:89   & $(1,-0.50, 0)$         &  $\dots$  & 0.0469  & 0.02516870 & 0.07199    & 0.01779  & $\dots$ &  0.14 \\
22 & SXS:BBH:1136 & $(1,-0.75,-0.75)$   &  $-1.90$ & 0.0777  &0.04288969 &0.12105      & 0.02728 & 0.074 & 0.10 \\
23 & SXS:BBH:321  & $(1.22,+0.33,-0.44)$& $+1.47$ & 0.0527  & 0.03239001 &0.07621     & 0.02694 & 0.015  &  0.21 \\
24 & SXS:BBH:322  & $(1.22,+0.33,-0.44)$& $-2.02$  & 0.0658  &  0.03396319 &0.0984       & 0.026895 & 0.016 & 0.20 \\
25 & SXS:BBH:323  & $(1.22,+0.33,-0.44)$& $-1.41$ & 0.1033  & 0.03498377 &0.1438      & 0.02584 & 0.019 & 0.15\\
26 & SXS:BBH:324  & $(1.22,+0.33,-0.44)$& $-0.04$ & 0.2018  & 0.02464165 &0.29421        & 0.01894 & 0.098 & 0.24\\
27 & SXS:BBH:1149 & $(3,+0.70,+0.60)$  &  $+3.00$  & 0.0371  &0.03535964 &$0.0621$   & $0.02664$ &0.025  & 0.96\\
28 & SXS:BBH:1169 & $(3,-0.70,-0.60)$    &  $+3.01$ & 0.0364  &0.02759632 &$0.04895$     & $0.024285$ & 0.033 & 0.10     %

 \end{tabular}
 \end{ruledtabular}
 \end{center}
 \end{table*}
\begin{figure*}[t]
\center
\includegraphics[width=0.31\textwidth]{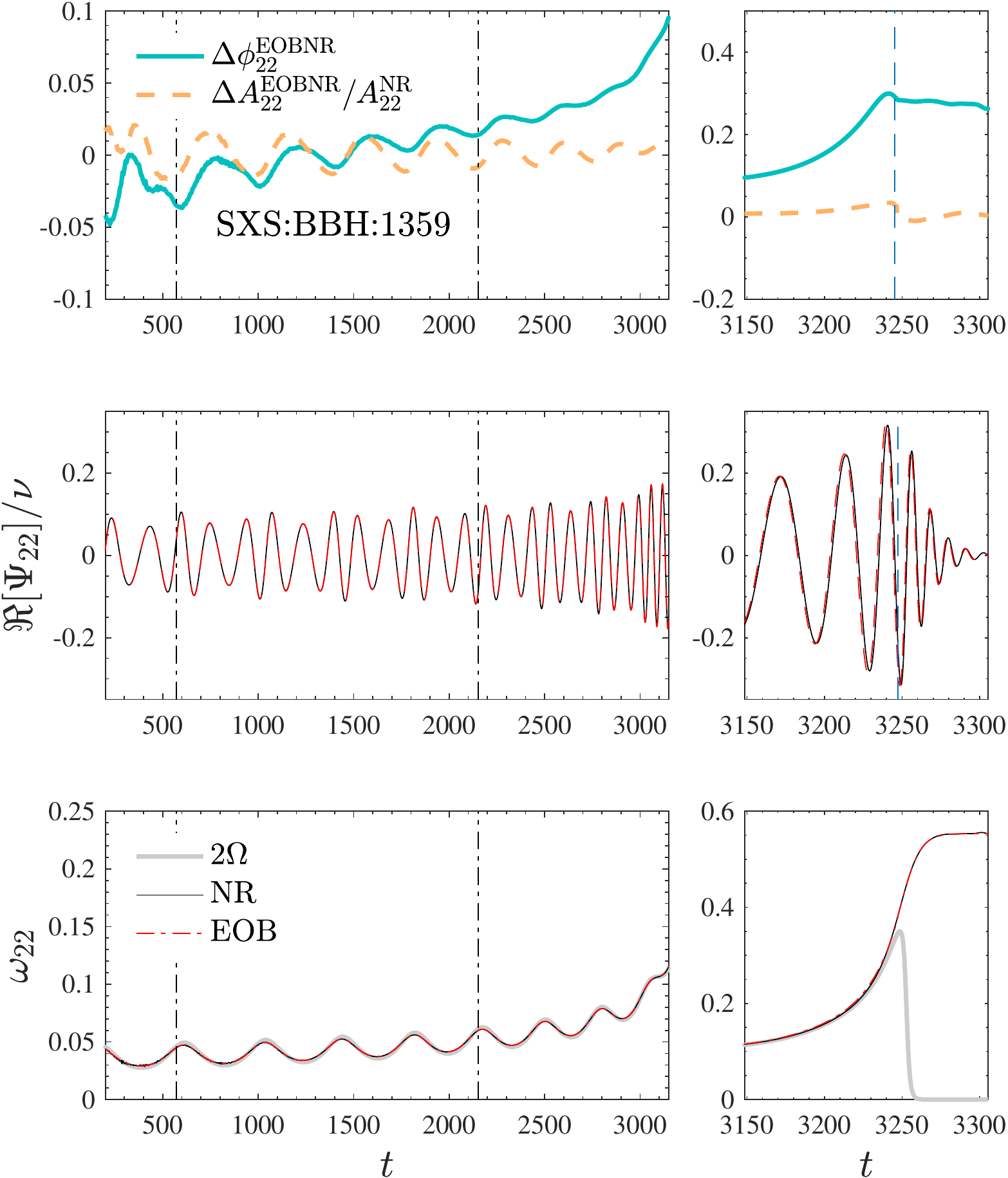} 
\includegraphics[width=0.31\textwidth]{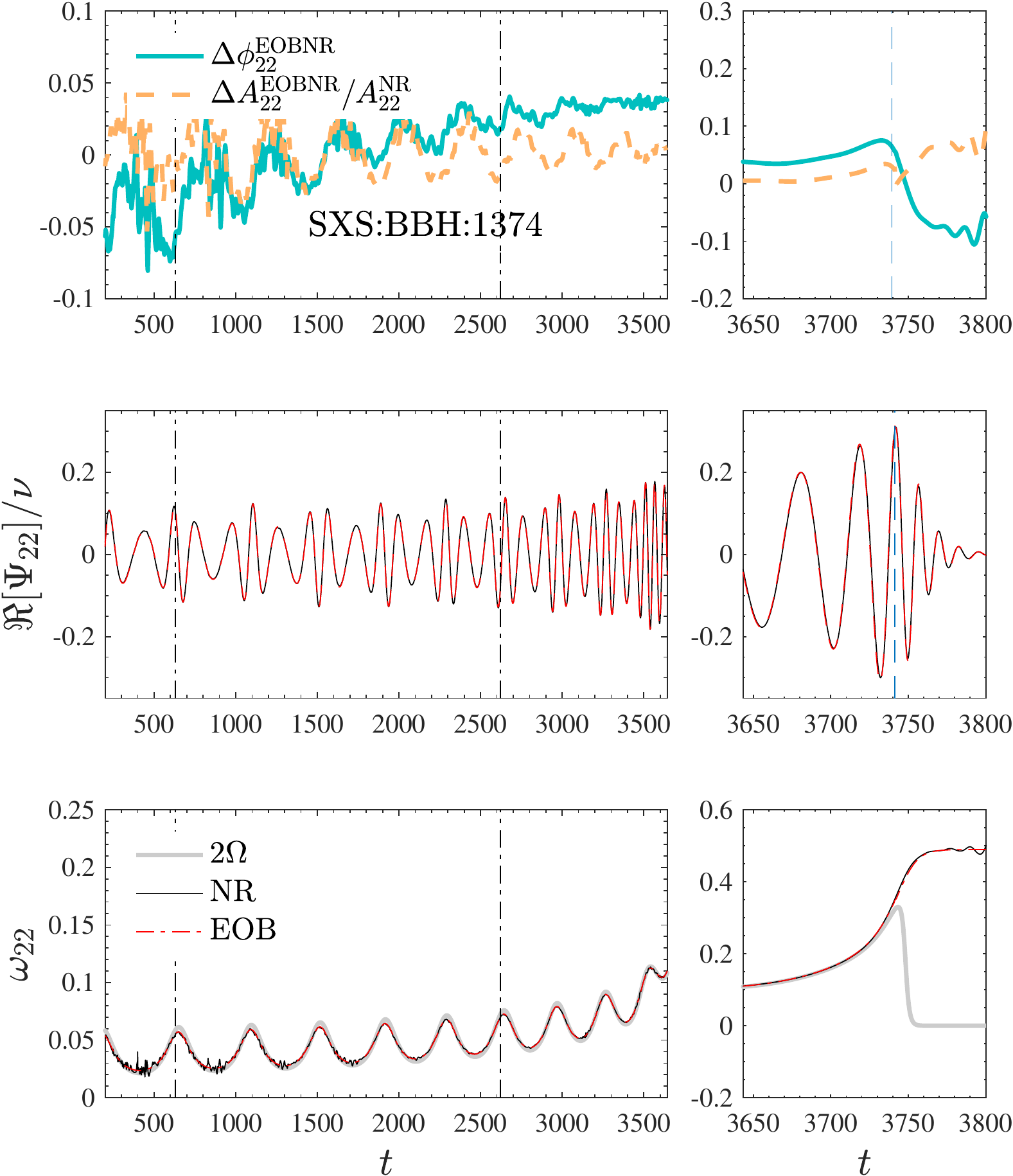}
\includegraphics[width=0.305\textwidth]{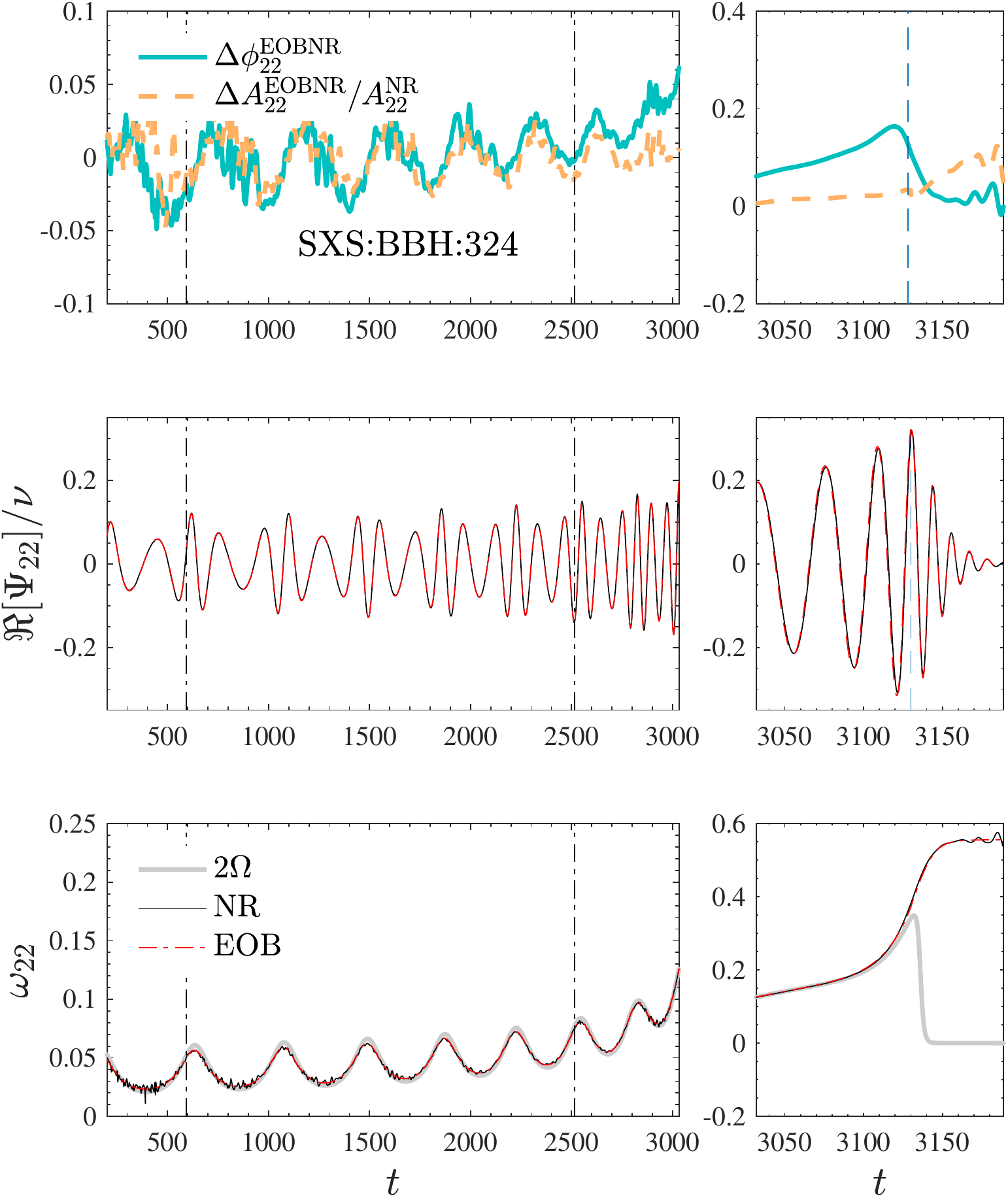}  
\caption{\label{fig:td_phasings}EOB/NR time-domain phasing comparison three meaningful configurations
SXS:BBH:1359, SXS:BBH:1374 and SXS:BBH:324. The phasing agreement is largely improved with respect
to the corresponding ones shown in Fig.~10 of Ref.~\cite{Nagar:2021gss}. The vertical dash-dotted lines in 
the left panels indicate the alignment interval, while the merger location is marked by a dashed vertical line
in the right panels.}
\end{figure*}

Let us finally evaluate the EOB/NR performance using the fully analytical expression for the N$^3$LO
spin-orbit contribution. First of all, Fig.~\ref{fig:q1_c40} displays time-domain phasing comparisons for
the same $q=1$ configurations considered above. The phase differences at merger are rather larger,
especially for large values of the individual spins.
The EOB/NR unfaithfulness computation is reported in Fig.~\ref{fig:barF_circ_cN3LO}: one finds that there
are many configuration even above the fiducial threshold of $3\%$. To have a simple understanding
of the inaccurate configurations it is helpful to plot $\bar{F}_{\rm EOB/NR}^{\rm max}$ versus the effective
spin $\tilde{a}_0$, Fig.~\ref{fig:max_barF}. One sees that the EOB/NR agreement degrades progressively
as the effective spin increases or decreases. In practice, the analytical
model can be considered robustly faithful ($<1\%$) only for mild values of the effective spin. Note however
that there is a region where $\bar{F}^{\rm max}_{\rm EOB/NR}< 1\%$ also for large, positive spins.
This corresponds roughly to simulations where $0.6 \lesssim \tilde{a}_0 \lesssim 0.8$ and $q < 5$.
For completeness, the same plot also reports (with green markers) the values of $\bar{F}^{\rm max}_{\rm EOB/NR}$
for the NR-informed value of $c_3$, so to give complementary information to the one of Fig.~\ref{fig:barF_circular}.
To conclude, what is striking in this comparison is that, similarly to the case of $a_6^c$ mentioned above a suitably
NR-tuned effective function is pragmatically {\it more efficient} than the outcome of a high-order analytical calculation.

In this respect, we recall that in our definitions of $(G_S,G_{S_*})$ we introduced two formal  N$^4$LO terms,
where $c_{40}^*= 2835/256$, fixed to the spinning test-mass value. However, analogously to the case of $c_3$,
we can {\it flex} these two coefficients as $c_{40}=\nu c_4 u_c^4$ and $c_{40}^*=\nu c_4 + 2835/256$ introducing
an effective N$^4$LO parameter that can be tuned to NR simulations analogously to $c_3$. We show this here explicitly
by determining $c_4$ for the specific case of equal-mass, equal spin binaries and evaluating the resulting performance
in terms of phasing.

Let us start by inspecting Fig.~\ref{fig:q1_c40}. The EOB/NR phase difference is always positive. However, the physical
meaning of this observation is different whether the spins are aligned or anti-aligned with the orbital angular momentum.
Let us start with the case $(1,+0.60,+0.60)$, middle right panel of  Fig.~\ref{fig:q1_c40}. The fact that the sign is positive
indicates that the transition from inspiral to plunge and merger occurs faster than the NR prediction, so 
to yield an accumulated phase difference of 1.5~rad at merger. 
Within the EOB Hamiltonian, this effect is interpreted as an underestimation of the spin-orbit interaction
with respect to the NR case. To contrast this fact, one should reduce the phase acceleration, or, in other words, 
increase the spin-orbit interaction so that its repulsive character (because spins are aligned with the orbital 
angular momentum) becomes larger. 
Similarly, for the case $(1,-0.60,-0.60)$ with $c_4=0$ one again finds that the EOB plunge phase is more 
accelerated than the corresponding NR one, but now the motivation is different: the spin-orbit interaction 
is now too large and it should be reduced.
To probe that $c_4$ can be tuned the same way as $c_3$, we consider a small set of equal-mass, equal-spin
configurations, see Table~\ref{tab:c4}. The table lists the first-guess values of $c_4$ that are found to yield
an acceptable ($\lesssim 1$~rad) phase difference at merger time. 
The interesting fact is that, analogously to the $c_3$ case, the magnitude of the effective parameter 
{\it increases} as the spins become large and negative. These $c_4$ data can be fitted by a quadratic
function of $\tilde{a}_0$, yielding the following function 
\be
\label{eq:c4}
c_4=39.43 \tilde{a}_0^2 - 141.77 \tilde{a}_0 + 65.73.
\ee
The corresponding time-domain comparison (either phase difference and amplitude difference) 
are shown in Fig.~\ref{fig:q1c4_tuned}. The phase difference at merger is notably 
reduced with respect to the $c_4=0$ case of Fig.~\ref{fig:q1_c40},  although it is still slightly 
less good than the simple NR-tuned $c_3$ case of Fig.~\ref{fig:q1c3_tuned}.
It seems thus that the use of the complete analytical N$^3$LO spin-orbit information
within the current model just moves the need of a NR-tuned parameter at the N$^4$LO order
with slightly less accuracy and with no real advantage. This suggests that, within the 
current model, it is more efficient to simply adopt the NR-tuned $c_3$ parameter.

\section{Eccentric inspiral configurations}
\label{sec:ecc_dynamics}
Let us move now to discussing eccentric inspirals. To do so, we precisely repeat 
here the analysis of Ref.~\cite{Nagar:2021gss}, see Secs.~IIIC and~IIID therein.
The initial conditions are slightly fine tuned with respect to Ref.~\cite{Nagar:2021gss},
so that Table~\ref{tab:SXS} is an updated version of Table~III of Ref.~\cite{Nagar:2021gss}.
All NR quantities are evidently the same. The EOB quantities are coming with updated computations
with the model discussed here, in particular one notices: (i) new initial conditions $(e_{\omega_a}^{\rm EOB},\omega_a^{\rm EOB})$
and (ii) new values of the maximum of the EOB/NR unfaithfulness $\bar{F}^{\rm max}_{\rm EOB/NR}$.
To start with, Fig.~\ref{fig:td_phasings} reports the $\ell=m=2$ time-domain phasing
 comparison for SXS:BBH:1359, SXS:BBH:1374 and SXS:BBH:324. For each configuration, (i) at the 
 top we have the phase difference and the relative amplitude difference; (ii) in the middle we compare
 the real parts of the waveform; (iii) in the bottom panel we compare the EOB and NR GW frequency,
 together with twice the orbital frequency $\Omega$. The phasing agreement is largely improved with 
 respect to what shown in Fig.~10 and Fig.~14 of Ref.~\cite{Nagar:2021gss}: the EOB/NR phase difference 
 is rather low and does not vary much during the inspiral and remains of the order of $0.1-0.2$ rad up
 to merger as well.
\begin{figure}[t]
	\center
	\includegraphics[width=0.44\textwidth]{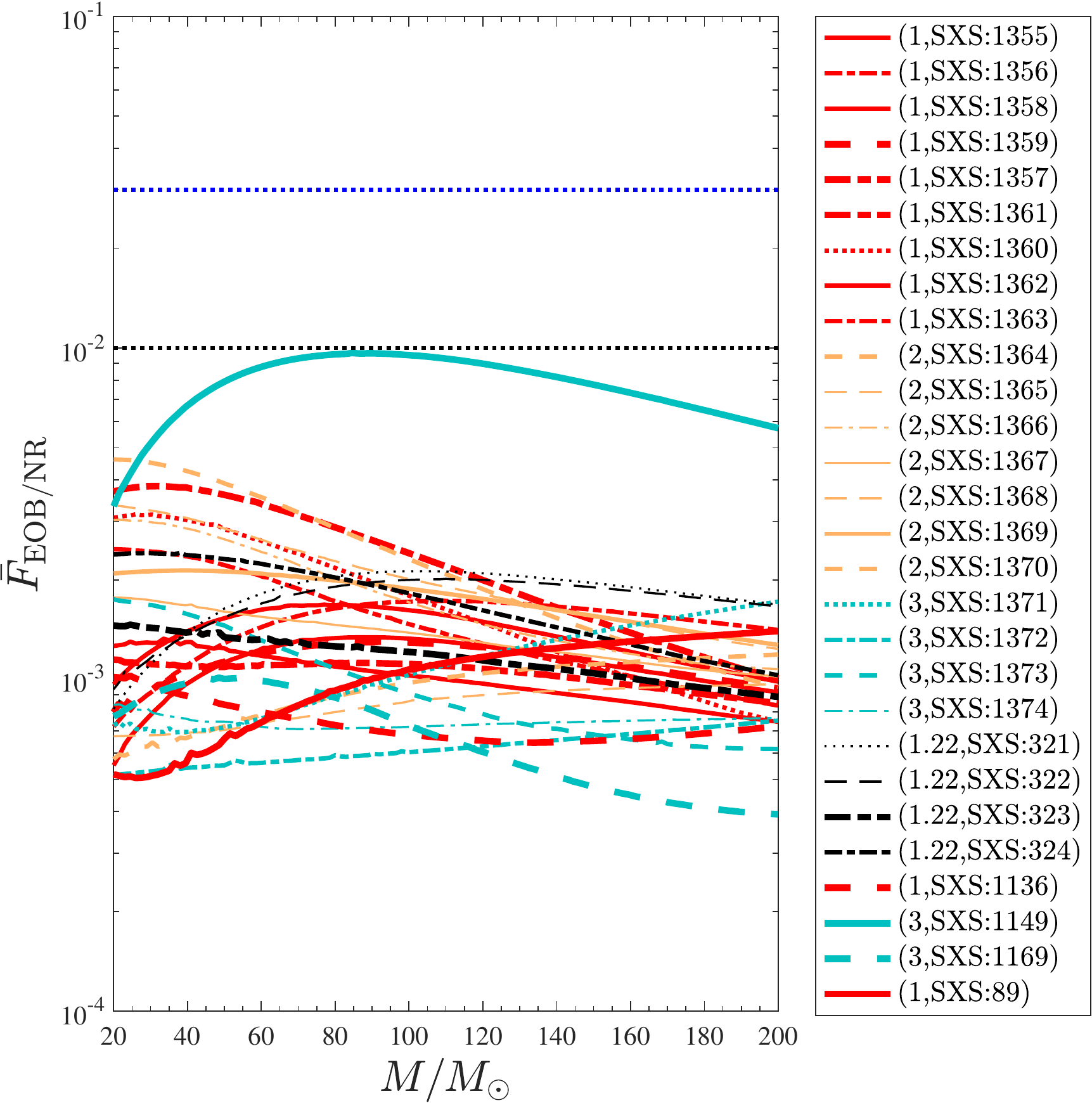} 
	\caption{\label{fig:barF_eccentric}EOB/NR unfaithfulness for the $\ell=m=2$ mode computed over the eccentric SXS 
	             simulations publicly available, Table~\ref{tab:SXS}. The horizontal lines mark the 0.03 and 0.01 values. 
		    All configurations are well below the $1\%$ except for SXS:BBH:1149, corresponding to $(3,+0.70,+0.60)$ 
		    with $e_{\omega_a}^{\rm NR}=0.037$, that is grazing this value. 
		    This is consistent with the slight degradation of the model performance for large positive spins, 
		    as found in the quasi-circular limit.}
\end{figure}

\begin{figure}[t]
	\center
	\includegraphics[width=0.4\textwidth]{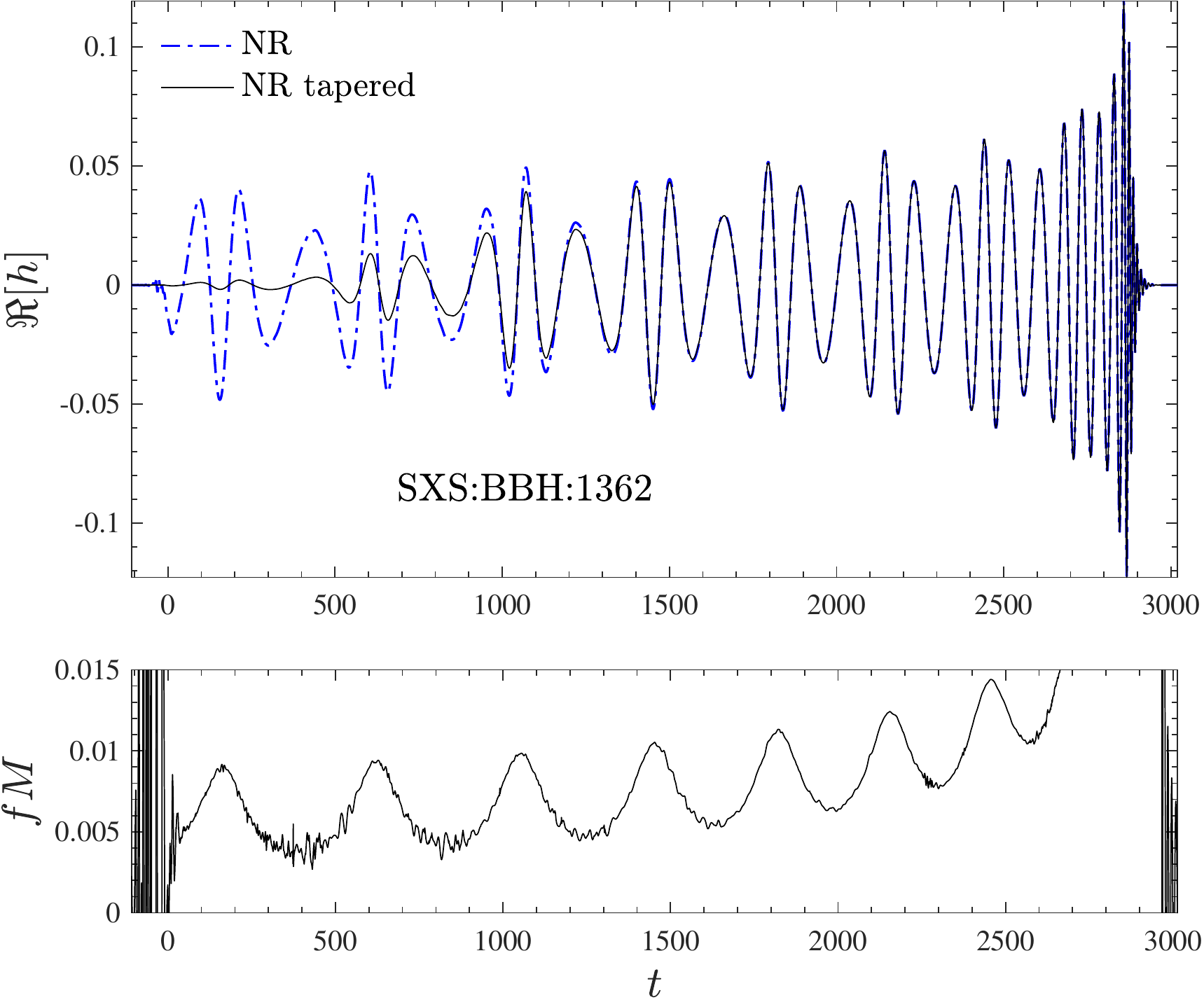} 
	\caption{\label{fig:1362_tapering}Illustrative example of the tapering procedure for SXS:BBH:1362. Note that the junk radiation
	in the NR dataset is practically absent.}
\end{figure}

\begin{figure}[t]
	\center
	\includegraphics[width=0.4\textwidth]{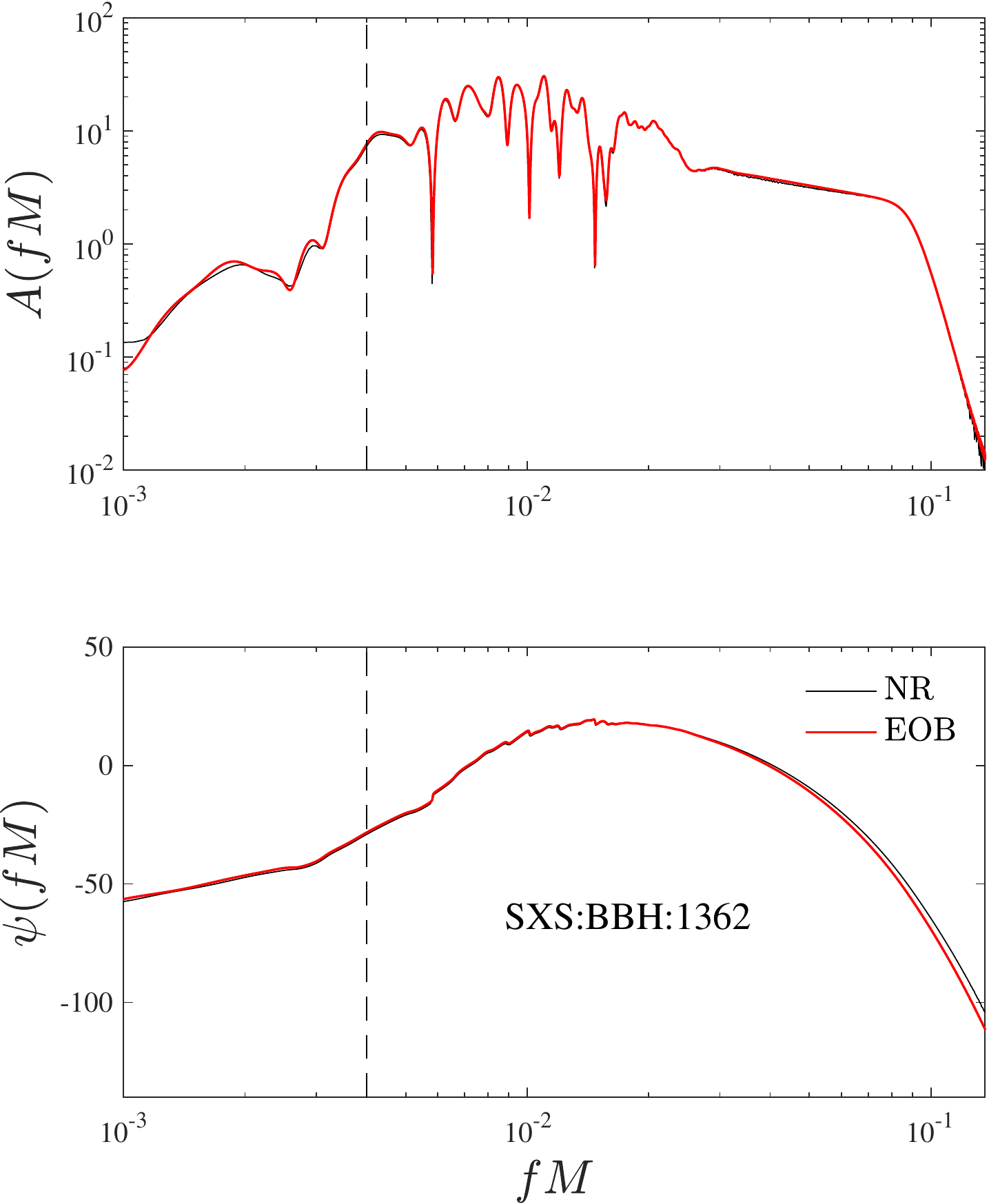} 
	\caption{\label{fig:1362}Comparing the NR (black) and EOB (red online) Fourier transforms for the SXS:BBH:1362 configuration: 
	amplitude ( $A(fM)$, top panel) and phase ($\psi(fM)$, bottom panel). The vertical dashed line indicates the lowest boundary of 
	integration to be used to compute the unfaithfulness from Eq.~\eqref{eq:barF}. Note that to accurately capture the oscillatory behavior
	of $A(fM)$ the time-domain vectors are padded with zeros before doing the Fourier transform. The part before the maximum of
	the amplitude (say $fM\lesssim 6\times 10^{-3}$, approximately corresponding to a kink in $\psi(Mf)$) has little influence 
	on $\bar{F}_{\rm EOB/NR}$ so that the initial frequency can be augmented up to this value without relevant consequences. 
	See text for additional discussion.}
\end{figure}

The global vision of the model performance is given by Fig.~\ref{fig:barF_eccentric}, that highlights the EOB/NR unfaithfulness
versus the total mass of the system. We find that all configurations are well below $1\%$ except for SXS:BBH:1149, that is grazing
this value. This is not surprising since SXS:BBH:1149 has parameter $(3,+0.7,+0.6)$, that give $\tilde{a}_0=0.675$,
a value that belongs to the region of $\tilde{a}_0$ where it is not possible to obtain a highly NR-faithful modelization 
already in the quasi-circular case. Despite this, the improvements in the quasi-circular sector reflect 
all over the $\bar{F}_{\rm EOB/NR}$ behavior of Fig.~\ref{fig:barF_eccentric}, either for small or 
for large eccentricities. This is in particular the case for the $q=1$ configurations, 
where $\bar{F}_{\rm EOB/NR}$ gets down to $\sim 10^{-3}$.
This is a remarkable improvement with respect to the results of Ref.~\cite{Nagar:2021gss}, 
where $\bar{F}_{\rm EOB/NR}$ was grazing the $1\%$ threshold for these configurations (see Fig.~11 therein).
In this respect, in redoing the $\bar{F}_{\rm EOB/NR}$ calculation for the current model, we
realized that Ref.~\cite{Nagar:2021gss}, overlooked the importance of the resolution employed in the calculation
of the Fourier transforms and its eventual impact on the results.  We thus revised and updated
also the $\bar{F}_{\rm EOB/NR}$ calculation of Ref.~\cite{Nagar:2021gss}, 
that is now discussed in Appendix~\ref{sec:barF_NBR}.
Entering the details, let us first recall that the NR signal is tapered before doing the Fourier transform so
as to reduce high-frequency spurious oscillations. Figure~\ref{fig:1362_tapering} illustrates this procedure
(obtained multiplying the waveform by an hyperbolic tangent function) on the NR dataset SXS:BBH:1362.
When one then takes the Fourier transform, we found it necessary (as already mentioned in Ref.~\cite{Hinder:2017sxy})
to pad the time-domain vectors with zeros in order to increase the frequency resolution and capture all the
details at low frequency.
The Fourier transform  of the NR signal is shown in Fig.~\ref{fig:1362} together with the one of the
corresponding EOB waveform (red online). The top-panel reports the amplitude $A(fM)$  and the bottom panel 
the phase $\psi(fM)$ [we assumed that $\tilde{h}=A(fM)e^{i\psi(fM)}$].
To accurately capture all the oscillatory behavior of the modulus, and also to avoid unphysical
features in the phase, we use here frequency resolutions $M\delta f\sim 10^{-4}$.
For each NR dataset we have been careful to progressively increase the resolution until the calculation of $\psi(fM)$
becomes robust. The same is done also for the EOB waveform, with the distinction that the corresponding tapering
should be chosen so to match as much as possible the behavior of the corresponding NR waveform.
Figure~\ref{fig:1362} refers to the SXS:BBH:1362 dataset, although the behavior is however typical for all other
configurations. The vertical line, located at $fM=0.004$, indicates the lower integration boundary, $Mf_{\rm min}$
in the unfaithfulness computation for this configuration. From Fig.~\ref{fig:1362_tapering} this value is close to
the minimal frequency reached at the first apastron. Clearly, the value of $Mf_{\rm min}$ may vary from one dataset 
to the other, although it is chosen to be always lower than the maximum of $A(fM)$. In any case, the part of the signal 
up to the amplitude maximum (say up to $fM\sim 0.006$ in this case) has negligible influence on the total result.
In obtaining the $\bar{F}_{\rm EOB/NR}$ results displayed in Fig.~\ref{fig:barF_eccentric} we have been careful
to obtain EOB/NR visual agreement between the Fourier transforms of each dataset analogous to the one shown
in Fig.~\ref{fig:1362} for the SXS:BBH:1362 dataset. 

Given the exploratory character of the current study, we have just briefly looked at higher modes. 
The NR-accurate behavior of 
all waveform modes during the inspiral is comparable to what discussed in Ref.~\cite{Nagar:2021gss}. 
By contrast, for what concerns merger and ringdown, although the modes with $m=\ell$ usually (though not always) look 
generally sane, those with $m\neq \ell$ may develop unphysical behaviors due to the action of NQC corrections, as already 
noted in Ref.~\cite{Nagar:2021gss}. This problem, that has always been present within \TEOBResumS{}~\cite{Nagar:2018zoe}, 
is now even amplified because of the existence of an effective horizon corresponding to the fact that the $A$ function
has a zero at a finite value of $u$. The issue of robustly determining NQC corrections for any multipole will require more dedicated 
investigations, that we will postpone to future work. We only anticipate that it is likely that a deeper understanding of 
NQC corrections (especially in relation with the dynamics) in the test-mass limit~\cite{Albanesi:2021rby} will be required 
to overcome what currently seems to be the most evident Achilles' heel of \TEOBResumS{}-based waveform models.

\section{Hyperbolic encounters and scattering angle}
\label{sec:scattering}
\begin{table*}[t]
 \caption{\label{tab:chi_scattering} Comparison between EOB and NR scattering angle. From left to right the columns report:
 the ordering number; the EOB impact parameter $r_{\rm min}$; the NR and EOB radiated energies, 
 $(\Delta E^{\rm NR}/M,\Delta E^{\rm EOB}/M)$; the NR and EOB radiated angular momentum, 
 $(\Delta J^{\rm NR}/M^2,\Delta J^{\rm EOB}/M^2)$; the NR and EOB scattering angles $(\chi^{\rm NR},\chi^{\rm EOB})$ and
 their fractional difference  $\hat{\Delta}\chi^{\rm NREOB}\equiv |\chi^{\rm NR}-\chi^{\rm EOB}|/\chi^{\rm NR}$.}
   \begin{center}
     \begin{ruledtabular}
\begin{tabular}{c c c c c c c c c } 
$\#$  & $r_{\rm min}$ & $\Delta E^{\rm NR}/M$ & $\Delta E^{\rm EOB}/M$ & $\Delta J^{\rm NR}/M^2$ & $\Delta J^{\rm EOB}/M^2$ & $\chi^{\rm NR}$ [deg] & $\chi^{\rm EOB}$[deg] & $\hat{\Delta}\chi^{\rm NREOB}[\%]$ \\
\hline
1 & 3.512    & 0.01946(17)       & 0.018785   &   0.17007(89)  & 0.159229    &   305.8(2.6)  & 317.349057  &   3.7767\\
2 & 3.79    & 0.01407(10)       & 0.012886  &  0.1380(14)     &  0.119728  & 253.0(1.4)   &  257.702771  &   1.8588\\
3 & 4.09    & 0.010734(75)     & 0.009449    &   0.1164(14)    &  0.095350  &  222.9(1.7)   & 224.412595  &   0.6786 \\
4 & 4.89    & 0.005644(38)     & 0.004612    & 0.076920(80)  &  0.057402  &  172.0(1.4)   & 171.283157  &   0.4168 \\
5 & 5.37    & 0.003995(27)     &  0.003175  & 0.06163(53)     &  0.044473  &152.0(1.3)   &  151.118180  &   0.5801\\
6 & 6.52     & 0.001980(13)    & 0.001524    & 0.04022(53)    &   0.027313  &  120.7(1.5)  & 119.931396  &   0.6368\\
7 & 7.61     & 0.0011337(90)  & 0.000867    &  0.029533(53) &   0.018971  &  101.6(1.7)  & 101.066199  &   0.5254\\
 8 & 8.68    & 0.007108(77)    & 0.000547   &  0.02325(47)    &  0.014158 & 88.3(1.8)    &   87.965443  &   0.3789\\
 9 &  9.74   &  0.0004753(75) & 0.000371   &  0.01914(76)    &  0.011084  &78.4(1.8)    & 78.168216  &   0.2956\\
10 & 10.79 & 0.0003338(77)  & 0.000265   &   0.0162(11)     &   0.008982  &  70.7(1.9)    &   70.477124    &  0.3152\\
  \end{tabular}
 \end{ruledtabular}
 \end{center}
 \end{table*}

To conclude, we present a new calculation of the EOB scattering angle from hyperbolic encounters 
and compare it with the few NR simulations available, updating the results of Ref.~\cite{Nagar:2020xsk,Nagar:2021gss}.
The changes in the conservative part of the dynamics will impact quantitatively on the 
scattering angle computation of Ref.~\cite{Nagar:2020xsk,Nagar:2021gss}, although the
phenomenology remains unchanged. We repeat here the EOB calculation  of the scattering angle 
$\chi$ for the 10 configurations simulated in NR~\cite{Damour:2014afa}
and that are discussed in Table~I of~\cite{Nagar:2020xsk}. The EOB outcome,
together with the original NR values, $(\chi^{\rm EOB},\chi^{\rm NR})$ is listed in Table~\ref{tab:chi_scattering}.
The table also reports the GW energy, $\Delta E$, and
angular momentum, $\Delta J$, losses for both the NR simulations and the EOB 
dynamics~\footnote{Let us specify that while the NR losses are computed from the waveform, the EOB
losses are computed subtracting the initial and final energy and angular momentum, i.e. effectively
accounting for the action of the radiation reaction on the dynamics.}. It is evident
the remarkable improvement with respect to the results of Ref.~\cite{Nagar:2021gss}. 
In particular, the strong-field configuration $\# 1$, shows an EOB/NR disagreement of only 
about $4\%$, {\it four times smaller} than the one Ref.~\cite{Nagar:2021gss}.
On top of validating the model for extreme orbital configurations, this finding is also 
a reliable cross check of the consistency and robustness of our procedure 
to obtain $a_6^c(\nu)$: although the function was determined using quasi-circular 
configurations, its impact looks to be essentially correct {\it also} for scattering configuration.
This makes us confident that our NR-informed analytical choices do represent a reliable,
though certainly effective, representation of the strong-field dynamics of two nonspinning
black holes.

\section{Conclusions}
\label{sec:conclusions}

We have explored the performance of a new EOB model for spin-aligned binaries for three
types of binary configurations: (i) quasi-circular inspiral; (ii) eccentric inspiral; (iii) hyperbolic scattering.
The novelty of this model is that it uses (or attempts to use) recently computed high-order PN
information in both the orbital and spin sector. Our findings are as follows:
\begin{enumerate}
\item[(i)] In the nonspinning case, the best resummation option to incorporate (some of) the currently available 
5PN information in the EOB potentials $(A,D,Q)$ consists in using diagonal and near diagonal Pad\'e approximants. 
In this case, the performance of the model in the quasi-circular limit is essentially equivalent to the standard 
quasi-circular version of \TEOBResumS{}~\cite{Nagar:2020pcj}. The current model is thus a step forward 
with respect to the quasi-circular nonspinning limit of the model of Ref.~\cite{Nagar:2021gss}, although it is 
still relying on a single function, $a_6^c(\nu)$ that is informed by NR simulations.
In particular, the successful construction of a NR faithful waveform model illustrates the synergy between 
different building blocks  that incorporate missing physics, and eventually yields a reduced, or at least simpler, 
impact of the NR-informed functions. This is in particular evident seen the current analytical simplicity of the 
effective function $a_6^c(\nu)$, that is representable using the quadratic function of $\nu$ given by Eq.~\eqref{eq:a6c_fit}.

\item[(ii)] Results in the spinning case are globally more faceted. First of all, differently from previous work, we incorporate
spin-spin effects up to NNLO, where the centrifugal radius is now written in a straightforward factorized and resummed form.
Within this paradigm, we have explored two options for the spin-orbit sector: (i) on the one hand, we follow the usual
\TEOBResumS{} paradigm and include an effective  N$^3$LO spin-orbit correction through a parameter $c_3$ that is 
informed by NR simulations; (ii) on the other hand, we exploit recent analytical results~\cite{Antonelli:2020aeb} 
that provided the complete analytical expression for this contribution. This latter can be additionally modified through
the inclusion of a N$^4$LO effective spin-orbit term.
In the case of the NR-informed $c_3$ it is possible to obtain a model that is NR faithful in the usual sense, 
with $\bar{F}_{\rm max}\lesssim 3\%$. One should note however that the performance worsens specifically when 
the mass ratio and the spins are large and positive. This is related to the fact that the dynamics is very sensitive to 
tiny variations of $c_3$ (e.g, of order unity) and slight imperfections in the global fit of $c_3$ when $a_0\gtrsim 0.5$ 
may end up in relevant phase differences around merger that show up as worsening of $\bar{F}_{\rm EOB/NR}$.

By contrast, when the analytically known N$^3$LO spin-orbit information is implemented,  the model remains acceptably 
faithful in a more limited range of $-0.4\lesssim \tilde{a}_0\lesssim +0.4$, although the EOB/NR phase difference at merger
can be as large as several radians. For the special equal-mass, equal-spin case, we have also shown that the N$^3$LO-accurate 
analytical spin-orbit sector can be flexed and improved using an effective N$^4$LO function $c_4$ that can be tuned to 
NR simulations likewise $c_3$. This allows to achieve an EOB/NR phasing agreement that is comparable to, although
slightly less good than, the one obtained with the NR-tuned $c_3$ alone.
This result suggests that, at least within the current analytical paradigm, pushing the spin-orbit information to the currently 
completely know analytical level doesn't seem to be essential. It is possible, however, that with a different choice for
the functional form of the Hamiltonian, notably the one that uses a different gauge and incorporates the full Hamiltonian 
of a spinning particle~\cite{Rettegno:2019tzh}, high-order terms might make a difference\footnote{Evidently, since both the
NNLO spin-spin and the N$^3$LO spin-orbit corrections to the Hamiltonian have not been obtained independently with other 
techniques, there is the possibility that they might be (partly) incorrect. There is thus an urgent need of additional 
analytical work to completely check the results of Refs.~\cite{Levi:2015ixa,Levi:2016ofk,Antonelli:2020aeb,Antonelli:2020ybz}.}. 
We hope to tackle this kind of study in detail in future work.

\item[(iii)]
The improvement in the quasi-circular sector of the model also reflects on the modelization of eccentric inspirals.
We performed a EOB/NR waveform comparison analogous to the one of Ref.~\cite{Nagar:2021gss} and
we found that a rather small EOB/NR phase difference is maintained up to merger, especially for the
nonspinning datasets considered. This entails EOB/NR unfaithfulness $\bar{F}_{\rm EOB/NR}(M)$ 
that are always below $1\%$ and actually the $\lesssim 0.3\%$ for most configurations. This finding mirrors 
the improvement achieved in the  model in the description of the late-inspiral and plunge phase with 
respect to Ref.~\cite{Nagar:2021gss}.

\item[(iv)]
For the hyperbolic scattering case, we repeat the EOB/NR comparison of the scattering angle performed in
previous work~\cite{Damour:2014afa,Nagar:2020xsk,Nagar:2021gss}. Remarkably, the joint action of the increased 
PN information, new Pad\'e resummation and NR-informed $a_6^c(\nu)$ (notably, to quasi-circular NR simulations) 
allows for a further improvement of the EOB/NR agreement of the scattering angles discussed in Ref.~\cite{Nagar:2021gss}.
This amounts to a EOB/NR disagreement of only $\sim 4\%$ for the dataset with the smallest impact parameter, a factor
of $4$ smaller than the result of Ref.~\cite{Nagar:2021gss}. This consistency between the various configurations seems 
to suggest that, at least in the nonspinning case, the combination of the various analytical ingredients entering the model 
can offer a reliable and robust representation of the general BBH dynamics. In particular, the model presented here
can be used to provide improved analyses of GW190521 under the hypothesis of having been generated by the 
hyperbolic capture of two black holes~\cite{Gamba:2021gap}.

\end{enumerate}

\begin{acknowledgments}
We are grateful to A.~Albertini for organizing some NR simulations for us, to G.~Riemenschneider 
for help with the unfaithfulness computations in the quasi-circular limit. We are also warmly thankful to R.~Gamba 
for a careful reading of the manuscript and for additional cross checks of the EOB/NR unfaithfulness  for elliptic inspirals.
Computations were performed on the Tullio cluster at INFN Turin.
\end{acknowledgments}

\appendix
\section{A fresher view at the results of Ref.~\cite{Nagar:2021gss}}
\label{sec:barF_NBR}

\subsection{Eccentric inspirals}
\label{se:nbr_ecc}
%
\begin{figure}[t]
	\center
	\includegraphics[width=0.44\textwidth]{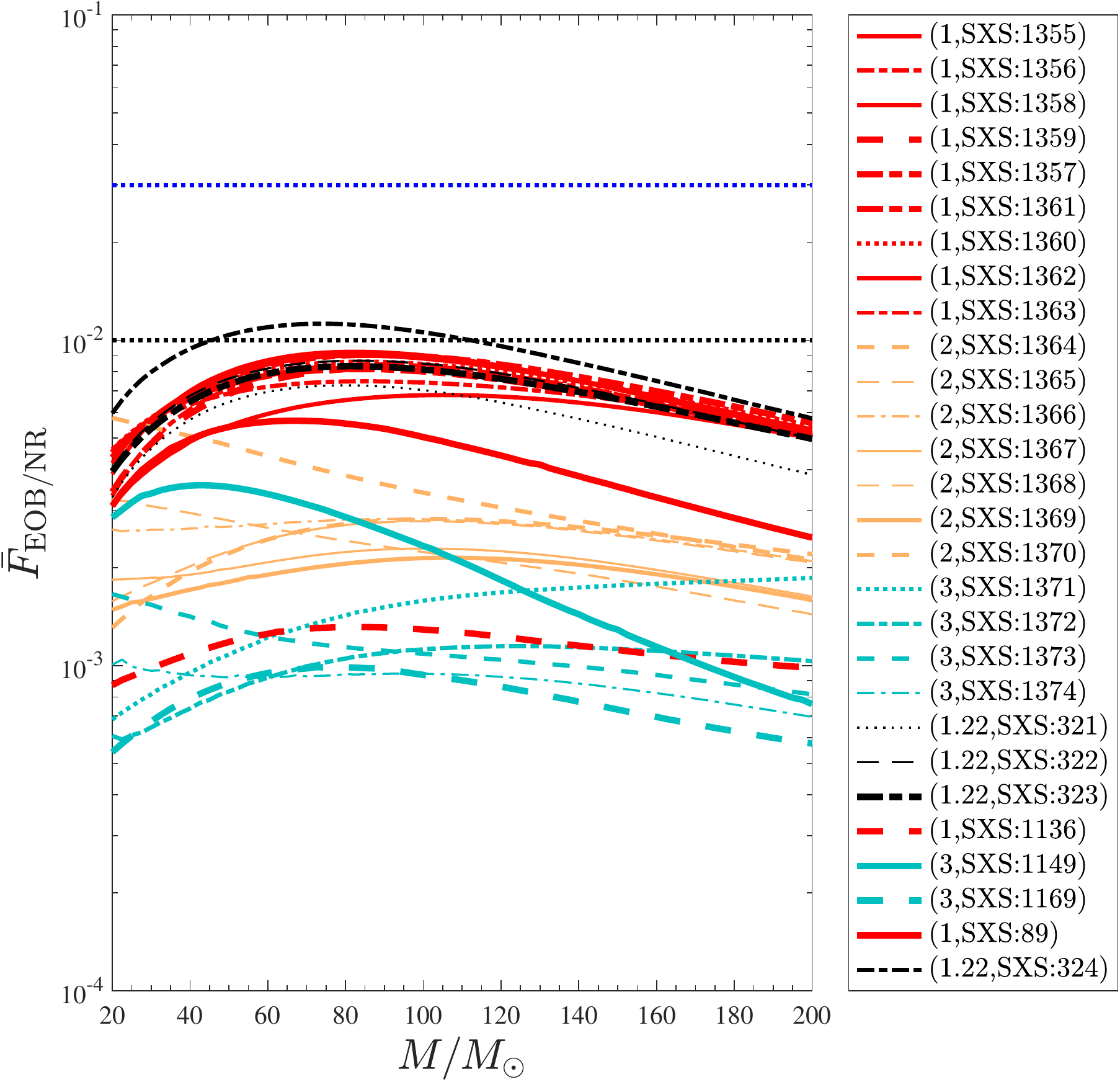} 
	\caption{\label{fig:barF_NBR}EOB/NR unfaithfulness for the $\ell=m=2$ mode computed over the eccentric SXS 
	             simulations publicly available using the model of Ref.~\cite{Nagar:2021gss} with a more accurate computation
	             of the Fourier transforms of both the EOB and NR waveforms. Comparing with Fig.~11 of Ref.~\cite{Nagar:2021gss}
	             one generally finds a better EOB/NR agreement  for $M\lesssim 80 M_\odot$.}
\end{figure}

\begin{table}[t]
   \caption{\label{tab:barF_NBR} Maximum values $\bar{F}_{\rm EOB/NR}^{\rm max}$ of the unfaithfulness shown in Fig.~\ref{fig:barF_NBR}.
   The third column of the table reports the lowest frequency used for integration, that approximately corresponds to the one of the second apastron
   of the NR waveform, so to cut the impact of the junk radiation.}
   \begin{center}
 \begin{ruledtabular}
   \begin{tabular}{cccc}
     $\#$ & ID & $M f_{\rm min}$ & $\bar{F}_{\rm EOB/NR}^{\rm max}[\%]$ \\
     \hline
1 & SXS:BBH:1355 & 0.0055 &0.92\\
2 &SXS:BBH:1356 &0.0057 &0.84\\
3 &SXS:BBH:1358 &0.006 & 0.91 \\
4 &SXS:BBH:1359 &0.0055 & 0.82 \\
5 &SXS:BBH:1357 &0.0055 &0.82 \\
6 &SXS:BBH:1361&0.0055 &0.90 \\
7 & SXS:BBH:1360&0.0055 &0.86 \\
8 & SXS:BBH:1362&0.006 &0.68 \\
9 & SXS:BBH:1363&0.006 &0.75 \\
10 & SXS:BBH:1364&0.0055 &0.3 \\
11 & SXS:BBH:1365&0.0055 &0.28 \\
12 & SXS:BBH:1366&0.0055 &0.28 \\
13 & SXS:BBH:1367&0.0055 &0.23 \\
14 & SXS:BBH:1368&0.0058 &0.32 \\
15 & SXS:BBH:1369&0.0056 &0.21 \\
16 & SXS:BBH:1370&0.0055 &0.58 \\
17 & SXS:BBH:1371&0.0054 &0.19 \\
18 & SXS:BBH:1372&0.0055 &0.11 \\
19 & SXS:BBH:1373&0.0055 &0.17 \\
20 & SXS:BBH:1374&0.0055 &0.10 \\
\hline
21 & SXS:BBH:89 &0.00465 & 0.56 \\
22 & SXS:BBH:1136 &0.0055 &0.13 \\
23 & SXS:BBH:321&0.0055 &0.73 \\
24 & SXS:BBH:322 &0.0055 &0.86 \\
25 & SXS:BBH:323 &0.0055 &0.83 \\
26 & SXS:BBH:324 &0.0058 &1.12 \\
27 & SXS:BBH:1149 &0.0054 & 0.36 \\
28 & SXS:BBH:1169 &0.004 & 0.099 
 \end{tabular}
 \end{ruledtabular}
 \end{center}
 \end{table}

 As pointed out in Section~\ref{sec:ecc_dynamics} of the main text, the EOB/NR unfaithfulness computation for
eccentric inspirals is sensitive to the resolution of the Fourier transforms, that should be high enough to
correctly resolve the low-frequency part of the (frequency-domain, FD) waveform. A FD resolution that is too
low may eventually fictitiously worsen the $\bar{F}_{\rm EOB/NR}$ values for small mass binaries (say $M\simeq 60M_\odot$) binaries.
This fact was unfortunately overlooked in previous works~\cite{Chiaramello:2020ehz,Nagar:2021gss}, and
it now corrected in this Appendix. 
 \begin{figure*}[t]
	\center
	\includegraphics[width=0.44\textwidth]{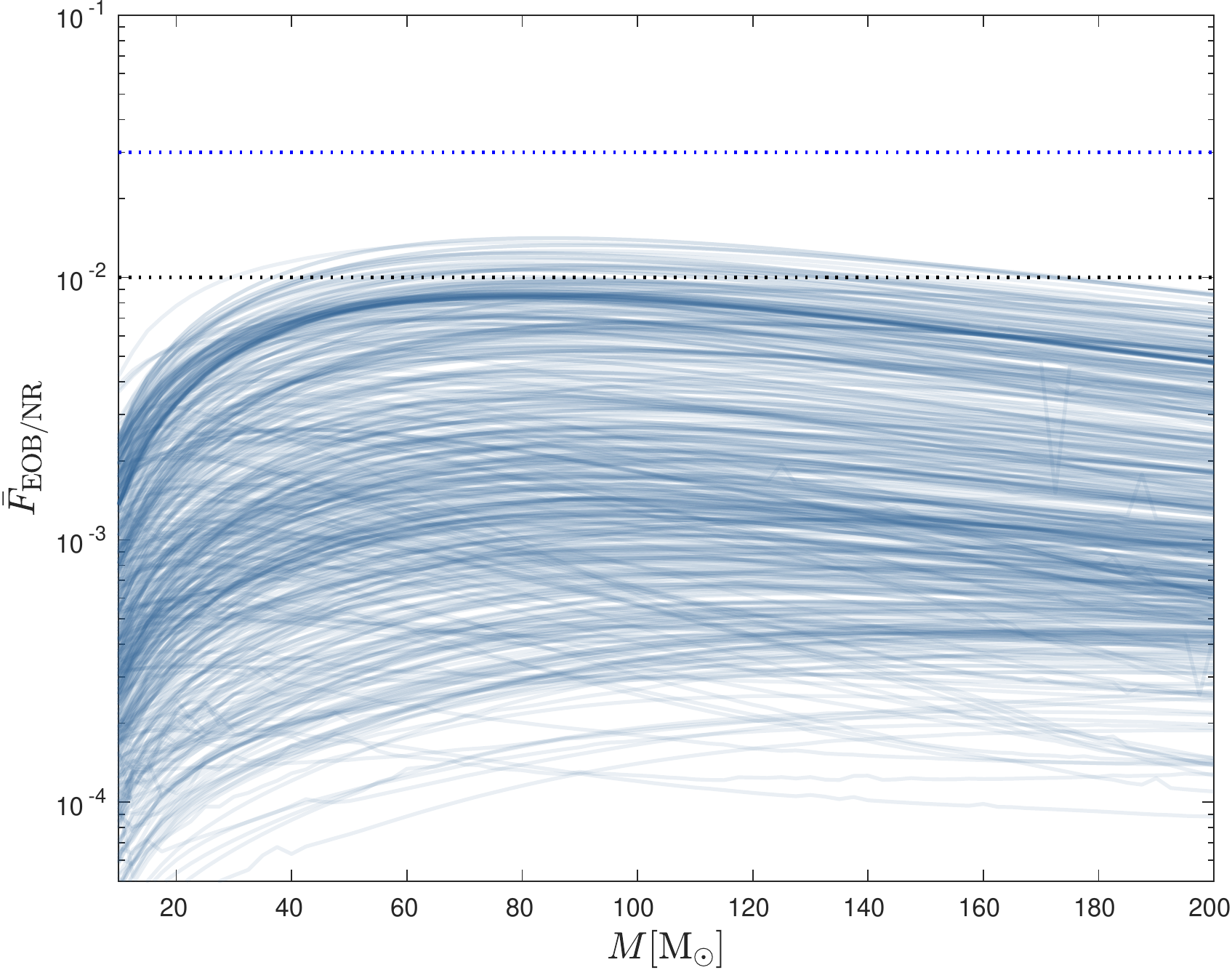}
	\includegraphics[width=0.44\textwidth]{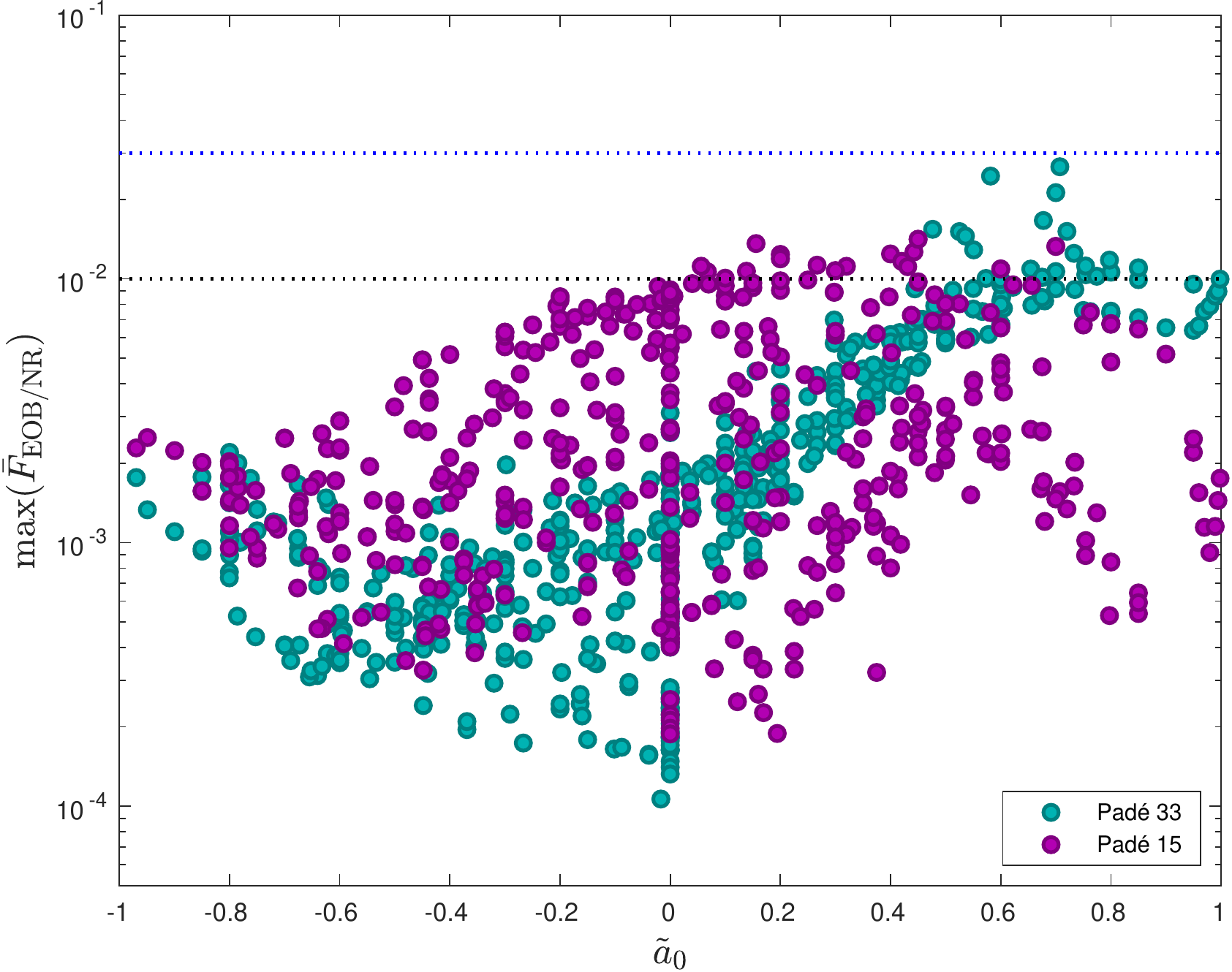} 
	\caption{\label{fig:barF_circ_NBR} 
		Left panel: unfaithfulness $\bar{F}_{\rm EOB/NR}(M)$ between the quasi-circular limit of the model 
		presented in Ref.~\cite{Nagar:2021gss} and the complete SXS catalog of noneccentric and 
		nonprecessing (spin-aligned) waveforms. Roight panel: global picture of the maximum EOB/NR unfaithfulness
		using the best model presented in this paper (Pad\'e 33, Fig.~\ref{fig:barF_circular}) and the one proposed in Ref.~\cite{Nagar:2021gss} 
		(Pad\'e 15, corresponding to the left panel of this figure). The black and blue dotted lines mark the $0.01$ and $0.03$ values respectively. 	
		The model of Ref.~\cite{Nagar:2021gss} is globally less NR faithful, but with a weak dependence on the effective spin parameter $\tilde{a}_0$.
		By contrast, the model used in this paper entails a very recognizable dependence on $\tilde{a}_0$.
		}
\end{figure*}
We focus then on the, more advanced, model of Ref.~\cite{Nagar:2021gss} and recomputed
the $\bar{F}_{\rm EOB/NR}$ quantity by: (i) increasing the resolution of the Fourier transforms,
obtained by suitably padding with zeros both the EOB and NR time-domain waveform; (ii) being careful 
to taper the EOB waveform so to have it visually consistent with the NR one and starting at approximately
the same frequency, like the case shown in Fig.~\ref{fig:1362}. 
Figure~\ref{fig:barF_NBR} is the updated version of Fig.~11 of Ref.~\cite{Nagar:2021gss}.
Qualitatively, the two results are largely consistent with the old ones, although the more accurate 
calculation of the Fourier transforms generally determines and improved EOB/NR agreement for
low values of $M$. In particular, it is interesting to note that for $M\sim 30 M_\odot$ one has 
$3\times 10^{-3}\lesssim\bar{F}_{\rm EOB/NR}\lesssim 5\times 10^{-3}$ for 
$q=1$ nonspinning binaries (with any initial eccentricity), that is about half the corresponding 
values reported in Fig.~11 of Ref.~\cite{Nagar:2021gss}. A similar improvement is found for all
other datasets, with the $\bar{F}_{\rm EOB/NR}$ curves that always mirror the improvement in
resolving the FD inspiral part of the waveform. However, Fig.~\ref{fig:barF_NBR} also shows that BBH:SXS:324 
remains somewhat an outlier, with the corresponding $\bar{F}_{\rm EOB/NR}$ value slightly above the $1\%$ level.
Some complementary information is also found in Table~\ref{tab:barF_NBR}, that reports the values of 
the maximal unfaithfulness $\bar{F}^{\rm max}_{\rm EOB/NR}$ together with the minimum values of 
the frequency $Mf_{\rm min}$ used in the calculation. For each dataset, these values approximately
correspond to the frequency of the second apastron of the NR waveform and are always smaller than 
the frequency corresponding to the maximum of the amplitude of the Fourier transform, as typically illustrated 
in Fig.~\ref{fig:1362} for the SXS:BBH:1362 dataset.

\subsection{Quasi-circular inspirals}
\label{sec:nbr_qc}
In the spirit of providing a clear comparison between the results of Ref.~\cite{Nagar:2021gss} and those discussed here, 
we also provide a new computation of $\bar{F}_{\rm EOB/NR}$ over the full SXS catalog of spin-aligned quasi-circular NR 
simulations  using the model of.~\cite{Nagar:2021gss} , analogously to what shown in 
Figs.~\ref{fig:barF_circular} and \ref{fig:barF_circ_cN3LO} in the main text. The result is shown in the
left panel of  Fig.~\ref{fig:barF_circ_NBR}, where we can see that the $\bar{F}^{\rm max}_{\rm EOB/NR}$ is mostly 
below $10^{-2}$, with only a few some outliers that are in any case well below $0.03$. This results complements and updates
the left panel of Fig.~3 of Ref.~\cite{Nagar:2021gss} that was considering, for simplicity, only a reduced, though significant,
sample of the full SXS catalog of 534 spin-aligned waveforms.
A clearer comparison with the model proposed in this paper, in particular Fig.~\ref{fig:barF_circular}, is shown in 
the right panel of Fig.~\ref{fig:barF_circular}, that contrasts the two values of $\bar{F}^{\rm max}_{\rm EOB/NR}$ 
versus the effective spin $\tilde{a}_0$ for all datasets available.

\bibliography{refs20211006.bib,local.bib}

\end{document}